\documentclass[preprint,pre,aps,amsfonts,amsmath,showpacs,superscriptaddress,nofootinbib]{revtex4}
\usepackage{amsmath,amssymb,amscd}
\usepackage{graphicx}
\usepackage{bm}
\usepackage{color}

\newcommand{\e}{{\rm e}}
\newcommand{\E}{{\mathbb E}}
\newcommand{\m}{\mathfrak{m}}

\newcommand{\bea}{\begin{eqnarray}}
\newcommand{\eea}{\end{eqnarray}}

\begin{document}
\title{Asymptotics for the Expected Maximum of Random Walks and L\'evy Flights with a Constant Drift}
\author{Philippe Mounaix}
\email{philippe.mounaix@cpht.polytechnique.fr}
\affiliation{Centre de Physique Th\'eorique, Ecole
Polytechnique, CNRS, Universit\'e Paris-Saclay, F-91128 Palaiseau, France.}
\author{Satya N. Majumdar}
\email{satya.majumdar@lptms.u-psud.fr}
\affiliation{LPTMS, CNRS, Univ. Paris-Sud, Universit\'e Paris-Saclay, 91405 Orsay, France.}
\author{Gr\'egory Schehr}
\email{gregory.schehr@lptms.u-psud.fr}
\affiliation{LPTMS, CNRS, Univ. Paris-Sud, Universit\'e Paris-Saclay, 91405 Orsay, France.}
\date{\today}
\begin{abstract}
In this paper, we study the large $n$ asymptotics of the expected maximum of an $n$-step random walk/L\'evy flight (characterized by a L\'evy index $1<\mu\leq 2$) on a line, in the presence of a constant drift $c$. For $0<\mu\leq 1$, the expected maximum is infinite, even for finite values of $n$. For  $1<\mu\leq 2$, we obtain all the non-vanishing terms in the asymptotic expansion of the expected maximum for large $n$. For $c<0$ and $\mu =2$, the expected maximum approaches a non-trivial constant as $n$ gets large, while for $1<\mu < 2$, it grows as a power law $\sim n^{2-\mu}$. For $c>0$, the asymptotic expansion of the expected maximum is simply related to the one for $c<0$ by adding to the latter the linear drift term $cn$, making the leading term grow linearly for large $n$, as expected. Finally, we derive a scaling form interpolating smoothly between the cases $c=0$ and $c\ne 0$. These results are borne out by numerical simulations in excellent agreement with our analytical predictions.
\end{abstract}
\pacs{05.40.Fb, 02.50.Cw}
\maketitle
%
%
\section{Introduction}\label{sec1}
Understanding the statistics of extremes of {\it correlated} random variables has lately led to an intense activity\ \cite{bouchaud_biroli,satya_review,third,review_record}. Consider a set of $n$ random variables $\{x_1, x_2, \cdots, x_n\}$ drawn from a joint distribution $P_{\rm joint}(x_1, x_2, \cdots, x_n)$ and let $M(n) = \max\{x_1, x_2, \cdots, x_n\}$ denote the maximum of this set. The generic extreme value problem consists in trying to find the statistics of $M(n)$, given the joint distribution $P_{\rm joint}(x_1, x_2, \cdots, x_n)$. In the case of independent and identically distributed (i.i.d.) random variables, each variable is chosen from the same distribution $p(x)$ and the joint distribution factorizes, $P_{\rm joint}(x_1, x_2, \cdots, x_n) = \prod_{i=1}^n p(x_i)$. This property substantially simplifies the study of $M(n)$ and its statistics is very well understood~\cite{gumbel}. On the other hand, if the variables are correlated, the joint distribution does not factorize and a general theory for computing the statistics of $M(n)$ is lacking, except for a few exactly solvable cases (for a brief review see \cite{Pal_Satya}). One example is a discrete-time random walk (RW) on a line, where the position $x_i$ of the walker after $i$ steps evolves via
\bea\label{def_rw_c0}
x_i = x_{i-1} + \eta_i ,
\eea
starting from $x_0 = 0$. The jump increments $\eta_i$'s are i.i.d. random variables, each drawn from a symmetric and piecewise continuous probability distribution function (PDF) $f(\eta)$. Note that even though the jump increments $\eta_i$'s are uncorrelated, the positions $x_i = \sum_{k=1}^i \eta_k$ are strongly correlated. In the following, we assume that the Fourier transform $\hat f(k) = \int_{-\infty}^{+\infty} e^{ik\,\eta}\,f(\eta) \, d\eta$ has the small $k$ behavior
\begin{equation}\label{eq1.3}
\hat f(k) = 1 - |a k|^\mu + O\left(|k|^{\nu}\right),
\end{equation}
where $a>0$ is the characteristic length scale of the jumps, $0 < \mu \leq 2$ is the L\'evy index $\mu$, and the subleading exponent $\nu > \mu$. 
For $\mu = 2$, the variance of the jump distribution $\sigma^2 = \int_{-\infty}^{+\infty} \eta^2 f(\eta) d\eta$ is finite and $a = \sigma /\sqrt{2}$. In this case, the suitably scaled RW converges to a Brownian motion as $n\rightarrow +\infty$. On the other hand, for $0 < \mu < 2$, $f(\eta)$ is a fat-tailed distribution, $f(\eta) \propto |\eta|^{-1-\mu}$ ($\eta \to \infty$), and the RW\ (\ref{def_rw_c0}) is a L\'evy flight of index $\mu$. Let $M(n) = \max\{x_0=0, x_1, \cdots, x_n\}$ be the maximum of the walk up to $n$ steps. One is then interested in the statistics of $M(n)$ for a generic jump PDF $f(\eta)$ whose Fourier transform has the small $k$ behaviour as in Eq. (\ref{eq1.3}). 

Computing directly the PDF of $M(n)$ for arbitrary $f(\eta)$ turns out to be highly non-trivial. However, there exists a somewhat indirect result
for the generating function of the Laplace transform of this PDF, known as the Pollaczeck-Spitzer formula\ \cite{Spi57} 
\bea\label{Poll_Spitzer_c0}
\sum_{n=0}^\infty s^n \, \E\left(\e^{-\lambda M(n)}\right) = \frac{1}{\sqrt{1-s}} \, \exp\left(-\frac{\lambda}{\pi} \int_0^{+\infty} \frac{\ln(1-s\,\hat f(k))}{\lambda^2 + k^2} \, dk\right) \;.
\eea
Although this formula is rather explicit, pulling out the asymptotics of the moments of $M(n)$ is far from straightforward. Even for the expected maximum, i.e. the first moment, the asymptotics for large $n$ are nontrivial~\cite{AS2005}. It turns out that this expected maximum, for any finite $n$, exists only for $1< \mu \leq  2$, while it is divergent for $0 < \mu \leq 1$.  Henceforth, we focus only on the case $1< \mu \leq  2$. The leading large $n$ behavior of $\E[M(n)]$ depends only on the exponent $\mu$ and is known to scale as $\sim n^{1/\mu}$ \cite{AS2005}. However, the next subleading correction of $\E[M(n)]$ depends on the exponent $\nu > \mu$ defined in Eq. (\ref{eq1.3}):  in the limit of large $n$, this correction term scales as $n^{1-(\nu-1)/\mu}$ for $\mu<\nu < \mu + 1$, while for $\nu > \mu +1$ it approaches a constant \cite{GLM17}. Indeed, in this latter case ($\nu > \mu +1$), one gets  
\begin{equation}\label{eqAS1}
\frac{\E [M(n)]}{a}=\frac{\mu\Gamma(1-1/\mu)}{\pi} n^{1/\mu}+\gamma
+O\left(\frac{1}{n^{1-1/\mu}}\right)\ \ \ \ \ (n\rightarrow +\infty),
\end{equation}
where $a$ is the jump length scale in Eq. (\ref{eq1.3}) and the subleading constant $\gamma$ is given explicitly by \cite{AS2005}
\begin{equation}\label{eqAS2}
\gamma =\frac{1}{\pi}\int_{0}^{+\infty}\ln\left(
\frac{1-\hat{f}(q/a)}{q^\mu}\right)
\, \frac{dq}{q^2} \;,
\end{equation}
where $\hat f(k)$ is the Fourier transform of the jump PDF. Note that, using the small $k$ behaviour of $\hat f(k)$ in Eq.\ (\ref{eq1.3}), it can be checked from Eq .(\ref{eqAS2}) that the constant $\gamma$ exists only for $\nu > \mu+1$. In the particular case of a uniform jump distribution over the interval $[-1,1]$, the value of this constant is relevant to the well known packing problem of random size rectangles in a two-dimensional strip \cite{AS2005,flajolet}. The result in (\ref{eqAS1}) has also been recently used in the context of calculating the mean perimeter of the convex hull for random walks in two-dimensions \cite{GLM17}.

In this paper, we generalize these results for the expected maximum to the case of a biased random walk with a non-zero drift $c$. We will restrict ourselves to the class of jump distributions defined in Eq.\ (\ref{eq1.3}) with $\nu\ge 2$ (and $1<\mu\le 2$). The evolution equation for the position of the walker in Eq. (\ref{def_rw_c0}) is now modified to
\begin{equation}\label{eq1.1}
x_i =x_{i-1}+c+\eta_i \;,
\end{equation}
with $x_0 =0$ and where the $\eta_i$'s are like in Eq.\ (\ref{def_rw_c0}). Let $M_c(n) = \max \{x_0=0, x_1, \cdots,x_n \}$ denote the maximum of the biased RW\ (\ref{eq1.1}) after $n$ steps. Like in the $c=0$ case, the generating function of the Laplace transform of the $M_c(n)$'s PDF with $c\ne 0$ can be written as a generalized version -- originally derived by Spitzer in\ \cite{Spi57} -- of the Pollaczeck-Spitzer formula (\ref{Poll_Spitzer_c0}). Unfortunately, the Spitzer's version is a rather long and cumbersome formula [see Eq. (\ref{Spit1}) of Appendix \ref{App_Spitzer}] which seems even harder to handle than in the $c=0$ case. In this paper, we put forward an alternative, equivalent, version of the Pollaczeck-Spitzer formula via generalizing to $c \neq 0$ an important result due to Hopf\ \cite{Hopf} (and independently to Ivanov\ \cite{Ivanov}) about half-space Green functions with $c=0$. The asymptotic large $n$ behavior of the expected maximum, $\E[M_c(n)]$, in the presence of a non-zero drift $c$ can then be extracted from this new version [see Eq. (\ref{eq3.5})]. This turns out to be much more convenient to use than the Spitzer's formula. We will see that the presence of a drift makes this asymptotic behavior highly non-trivial and interesting.

To appreciate what makes the large $n$ behavior of $\E[M_c(n)]$ for discrete time random walks different, it may be instructive to put it into perspective by first recalling the corresponding large $t$ behavior of $\E[M_c(t)]$ for a biased continuous time Brownian motion evolving as
\begin{equation*}
\frac{dx(t)}{dt}= c + \sqrt{2D}\, \eta(t)
\end{equation*}
where $c$ is the drift, $D$ is the diffusion constant, and $\eta(t)$ is a Gaussian white noise with zero mean and delta correlator, $\langle \eta(t)\eta(t')\rangle = \delta(t-t')$. We recall that, by virtue of the central limit theorem, Brownian motion and discrete time random walks with $\mu =2$ (i.e. jumps with a finite variance) are asymptotically equivalent at large time. For a Brownian motion without drift ($c=0$) it is a well known classical result that \cite{Yor_book} (for a simple derivation, see section \ref{sec:BM})
\begin{eqnarray}\label{BM_c=0}
\E[M(t)] = \frac{2}{\sqrt{\pi}} \sqrt{D \, t} \;,
\end{eqnarray}
For $c \neq 0$, the expected maximum $\E[M_c(t)]$ behaves quite differently. Indeed, for large $t$, one finds (see section \ref{sec:BM} for a simple derivation)
\begin{eqnarray}\label{BM_cneq0}
\E[M_{c \neq 0}(t)] \sim \theta(c)\, c\, t + \frac{D}{|c|}\ \ \ \ \ (t \to \infty) \;,
\end{eqnarray}
where $\theta(c)$ is Heaviside step function [$\theta(c) = 1$ if $c>0$ and $\theta(c) = 0$ if $c < 0$]. Note that for $c<0$, the expected maximum approaches a positive constant, $\E[M_c(t)]\to D/|c|$, as $t \to \infty$. This behavior can be qualitatively understood as follows: for $c<0$, the particle drifts towards $-\infty$ as $t$ gets large, with occasional finite excursions to the positive side at early times only. The maximum is thus mainly set by these early time positive excursions without being significantly affected by the late time part of the trajectory, hence the expected maximum tends to a positive constant as $t$ gets large.

It is also interesting to notice that Eqs. (\ref{BM_c=0}) and (\ref{BM_cneq0}) imply that the limit $c \to 0$ in the large $t$ asymptotic behavior is actually singular, in the sense that the limits $c \to 0$ and $t \to \infty$ do not commute. In fact, we show explicitly in section \ref{sec:BM} that, for any $c$ and $t$, the expected maximum for the Brownian motion is a function of the scaling variable $u = |c| \sqrt{t/D}$
\begin{equation}\label{scaling_mu2_1}
\E[M_c(t)] - \theta(c)\, c\, t = \sqrt{D\,t} \; {\cal G}_2\,\left(\frac{|c|\,\sqrt{t}}{\sqrt{D}} \right) \;,
\end{equation}
with
\begin{equation}\label{scaling_mu2}
{\cal G}_2(u) = \frac{\e^{-\frac{u^2}{4}}}{\sqrt{\pi}} + \frac{{\rm erf}{\left(\frac{u}{2}\right)}}{u} - \frac{1}{2} u\, {\rm erfc}\left(\frac{u}{2}\right) \;,
\end{equation}
where ${\rm  erf}(x) = \frac{2}{\sqrt{\pi}} \int_0^x e^{-t^2} \, dt$ and ${\rm  erfc}(x) =  1- {\rm erf}(x)$. The scaling function ${\cal G}_2(u)$ behaves as ${\cal G}_2(u) \sim 2/\sqrt{\pi}$ for $u \to 0$ (which corresponds to $c\to 0$ first, then $t\to\infty$) and ${\cal G}_2(u) \sim 1/u$ as $u \to \infty$ (which corresponds to $t\to\infty$ first, then $c\to 0$), thus interpolating smoothly between the behaviors given in Eqs. (\ref{BM_c=0}) and (\ref{BM_cneq0}), respectively.
 
Having recalled the large time behaviors of the expected maximum for a Brownian motion, we now summarize the main asymptotic results obtained in this paper for discrete time random walks with a drift $c$, as defined in\ (\ref{eq1.1}). Below, we will express our results in units of the characteristic length scale $a$ [see Eq. (\ref{eq1.3})].  

\begin{itemize}
\item [$\bullet$]{For $\mu=2$, we find that the large $n$ behavior of the expected maximum is given by
\begin{equation}\label{eq1.6a}
\frac{\E[M_c(n)]}{a}\sim\frac{c}{a} \theta(c)\, n + \frac{|c|}{a}\,\kappa_c\ \ \ \ \ (t \to \infty) \;,
\end{equation}
with
\begin{equation}\label{eq1.6b}
\kappa_c =\frac{1}{2\pi}\frac{\partial}{\partial\lambda}\left.\int_{-\infty}^{+\infty}
\frac{\ln\lbrack 1-\hat{f}(q/c){\rm e}^{-iq}\rbrack}{\lambda +iq}\, dq\right\vert_{\lambda =0} \;.
\end{equation}
In principle, Eq.\ (\ref{eq1.6b}) makes it possible to compute $\kappa_c$ for any jump distribution $f(\eta)$ with a finite $\sigma$, although the task may be arduous. For instance, for a Gaussian jump distribution, $f(\eta) = 1/(\sigma \sqrt{2 \pi})\,\e^{-\eta^2/(2 \sigma^2)}$, we get $\kappa_c$ as the series
\begin{eqnarray}\label{gamma_gaussian}
\kappa_c = \sum_{m=1}^\infty \left[\frac{\e^{-b^2\,m}}{2b \sqrt{\pi\,m}}  - \frac{1}{2} {\rm erfc}\left(b\sqrt{m} \right) \right] \;,
\end{eqnarray}
where $b=|c|/\sigma\sqrt{2}=|c|/2a$ and $a = \sigma/\sqrt{2}$ (see Sec. \ref{sec5.3} for details). The apparent similarity between the large $n$ behavior\ (\ref{eq1.6a}) and the large $t$ behavior\ (\ref{BM_cneq0}) is in agreement with the expected convergence of the RW to a biased Brownian motion as $n$ gets large. Nevertheless, the constant $|c|\kappa_c/a$ in Eq. (\ref{eq1.6a}) does depend on the jump distribution explicitly, unlike its counterpart in Eq.\ (\ref{BM_cneq0}). For $c<0$, this discrepancy can be explained by the fact mentioned before [see below Eq.\ (\ref{BM_cneq0})] that the maximum is determined by early time positive excursions occurring in a finite time interval (with probability one) from $t=0$. For such early time excursions, boundary effects -- sensitive to the discrete or continuous nature of time -- are expected to play a role, hence the discrepancy between the constant terms in Eqs.\ (\ref{eq1.6a}) and\ (\ref{BM_cneq0}). For $c>0$, the same reasoning holds where the starting point of the RW is replaced with its arrival point and the trajectory is considered as running backward in time (see Fig. \ref{fig_identity} below and the corresponding explanation in the text). As already observed in the Brownian case, the expression of $\E[M_c(n)]$ for $c=0$ in Eq. (\ref{eqAS1}) (with $\mu=2$) and the one for $c \neq 0$ in Eq. (\ref{eq1.6a}) show that the limit $c \to 0$ in the large $n$ asymptotic behavior is actually singular in the discrete time case too. As before [see Eq. (\ref{scaling_mu2_1})], we find that there is a scaling regime for $n \to \infty$, $c \to 0$ keeping the ratio $u = |c| \sqrt{n}/a$ fixed which interpolates between these two behaviors in Eqs. (\ref{eqAS1}) (with $\mu = 2$) and (\ref{eq1.6a}). In this scaling regime, one finds that $\E[M_c(n)]$ is described by the universal scaling form
\begin{eqnarray}\label{scaling_mu2_RW}
\frac{\E[M_c(n)] - c \, \theta(c)n}{a}\sim\sqrt{n} \, {\cal G}_2\left(\frac{|c|\sqrt{n}}{a}\right) \;,
\end{eqnarray}
where ${\cal G}_2(u)$ is given in Eq. (\ref{scaling_mu2}) and is thus independent of the jump distribution~$f(\eta)$. Note that from Eq.\ (\ref{eq1.6a}) and the large $|c|\sqrt{n}/a$ behavior of\ (\ref{scaling_mu2_RW}), $\kappa_c$ must diverge like $\kappa_c\sim a^2/|c|^2$ as $c\to 0$, for any jump distribution (with a finite $\sigma$). This is actually the case, as we will show at the end of Sec.\ \ref{sec5.3}.
}
\item[$\bullet$] For $1<\mu<2$, the discrete time RW converges in the large $n$ limit to a biased L\'evy process and one expects a qualitatively different behavior, compared to the Brownian motion. For large $n$, all the terms surviving the large $n$ limit in the large $n$ expansion of $\E[M_c(n)]$ can be computed explicitly [see Eq. (\ref{eq5.1.2.8}) below]. The two leading terms are given by
\begin{equation}\label{asym1}
\frac{\E[M_c(n)]}{a}\sim\frac{c}{a} \theta(c)\,n +  \frac{|c|}{a}\frac{C \,n^{2-\mu}}{2-\mu}\ \ \ \ \ (n\to\infty) \;,
\end{equation}
where 
\begin{equation}\label{const1}
C=\frac{\Gamma(\mu -1)}{\pi}\sin\left(\frac{\pi\mu}{2}\right)\left(\frac{a}{\vert c\vert}\right)^\mu \;.
\end{equation}
Surprisingly, it follows from Eq.\ (\ref{asym1}) that for $1 < \mu < 2 $ and $c<0$, $\E[M_c(n)]$ still grows with $n$, unlike in the $\mu=2$ case where it approaches a constant. This is somewhat unexpected since, in this case, one can write the process $x_n$ in Eq. (\ref{eq1.1}) as $x_n = y_n + c \,n $ where $y_n$ converges to a symmetric L\'evy flight for large $n$. This implies in particular that, typically, $y_n = O(n^{1/\mu})$ for $n \gg 1$, which is thus much smaller than the drift term $c \,n$. Hence, for $c<0$, although the walker will typically drift to $-\infty$, she/he will always perform rare big jumps that will contribute to higher and higher values of $\E[M_c(n)]$ as $n$ increases. A scaling form generalizing to $1<\mu <2$ the one in\ (\ref{scaling_mu2_RW}) for $\mu =2$ can be obtained in the limit $c \to 0$ and $n \to \infty$ keeping the product $u = |c| n^{1/\mu}/a$ fixed. One finds
\begin{equation}\label{scaling_mu_1}
\frac{\E[M_c(n)] - c \, \theta(c)n}{a} \sim n^{1/\mu} {\cal G}_\mu\left(\frac{|c|}{a} n^{1-1/\mu} \right) \:,
\end{equation}
with
\begin{equation}\label{scaling_mu}
{\cal G}_\mu(u) = \frac{\mu}{\mu-1} u^{-\frac{1}{\mu-1}} \int_0^u dy \, y^{\frac{2-\mu}{\mu-1}} \int_0^{+\infty} dz \, z\, \, f_{S,\mu}(z+y) \;,
\end{equation}
where $f_{S, \mu}(x)$ is the stable law of index $\mu$, i.e., $f_{S,\mu}(x) = \int_{-\infty}^{+\infty} \frac{dq}{2 \pi} \, \e^{-|q|^\mu - i qx}$. For $\mu=2$, it can be checked that the integral representation (\ref{scaling_mu}) coincides with the expression of ${\cal G}_2(u)$ given in Eq.~(\ref{scaling_mu2}). For generic $1 < \mu <2$, there is no explicit expression for ${\cal G}_\mu(u)$ but it can be plotted and one can also easily obtain its asymptotic behaviors. One finds ${\cal G}_{\mu}(u) \sim \mu\,\Gamma(1-1/\mu)/\pi$ as $u \to 0$  and ${\cal G}_{\mu}(u) \sim \Gamma(\mu-1)/(\pi(2-\mu)) \sin(\pi \mu/2)\, u^{1-\mu}$ for $u \to \infty$, thus ensuring a smooth matching with Eq. (\ref{eqAS1}) for $u \to 0$ and with Eq.~(\ref{asym1}) for $u \to \infty$.

\item[$\bullet$]{Finally, we notice an interesting identity in law for the maximum of a random walk in the presence of a non-zero drift $c$. Consider a realization of the first $n$ steps of the walk from $x_0 = 0$ to some $x_n$ at step $n$. As before,  let $M_c(n)$ denote the maximum of the trajectory for this realization (see Fig. \ref{fig_identity}). Then, the following identity in law holds
\bea\label{identity_txt}
M_c(n) - x_n \equiv M_{-c}(n) \;,
\eea
where '$\equiv$' means that the random variable on the left hand side has the same distribution as the random variable on the right hand side. This identity can be understood very simply with the help of the figure \ref{fig_identity}. 
\begin{figure}
\includegraphics[width = 0.8\linewidth]{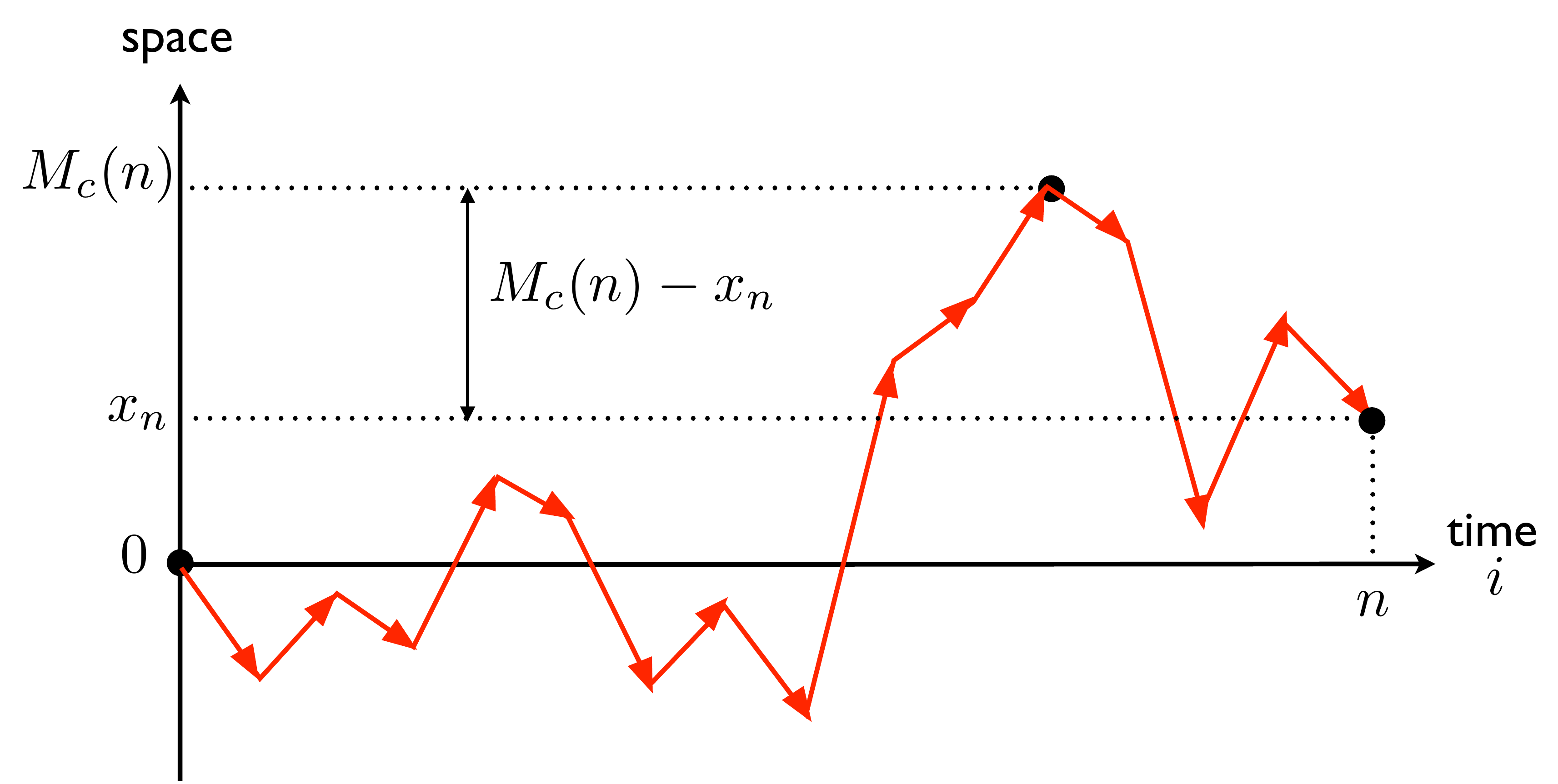}
\caption{Ilustration of the identity in law $M_c(n) - x_n \equiv M_{-c}(n)$ (see text and Eq. (\ref{identity_txt})).}\label{fig_identity}
\end{figure}
Consider the trajectory as running backward in time, starting at $x_n$, and set the new origin of space at $x_n$. Then, clearly, $M_c(n) - x_n$ represents the maximum of this shifted trajectory running backward in time. However, this backward process also corresponds to a random walk starting at the new origin and subjected to an opposite drift $-c$. Thus the maximum of this backward process is also a realization of $M_{-c}(n)$, which establishes the identity in law in Eq. (\ref{identity_txt}). As a consequence, taking expectation of both sides of Eq. (\ref{identity_txt}) and using $\E(x_n) = c\,n$, we get
\bea\label{identity_av}
\E[M_c(n)] - c\,n = \E[M_{-c}(n)] \;.
\eea   
One can easily check that our results in Eqs. (\ref{eq1.6a}) and (\ref{asym1}) are fully consistent with the general identity in Eq.~(\ref{identity_av}). 
}
\end{itemize} 

The rest of the paper is organised as follows. In Section \ref{sec:BM}, we discuss the statistics of the maximum $M_c(t)$ for the 
continuous time Brownian motion with a drift $c$. In Section \ref{sec2} we develop the formalism and the analytical tools needed to derive the asymptotic
behaviors of the expected maximum for a discrete time random walk with a non-zero drift. Specifically, we first generalize a formula by
Hopf\ \cite{Hopf} (and independently by Ivanov\ \cite{Ivanov}) for the driftless case to a non-zero drift. Then, we use this generalized formula to provide an alternative 
derivation of the Spitzer's formula\ \cite{Spi57} leading to a new, equivalent, version of the Spitzer's result. This new version proves to be much more convenient to extract the asymptotic behaviors for large $n$. Sections \ref{sec4} and \ref{sec5} contain the derivation of the main asymptotic results for the expected maximum, using the formalism developed in the previous section. In Section \ref{sec4}, we provide a re-derivation of the known results for the driftless case using this new formalism. We derive the detailed results for the case with a non-zero drift in Section \ref{sec5}. In Section \ref{sec6}, we verify our analytical predictions via numerical simulations. Finally, we conclude in Section \ref{sec7}. Some details are relegated to the Appendices. 
%
%
\section{Expected maximum of a $\bm{1D}$ Brownian motion in the presence of a 
constant drift}\label{sec:BM}

\begin{figure}
\includegraphics[width=0.7\linewidth]{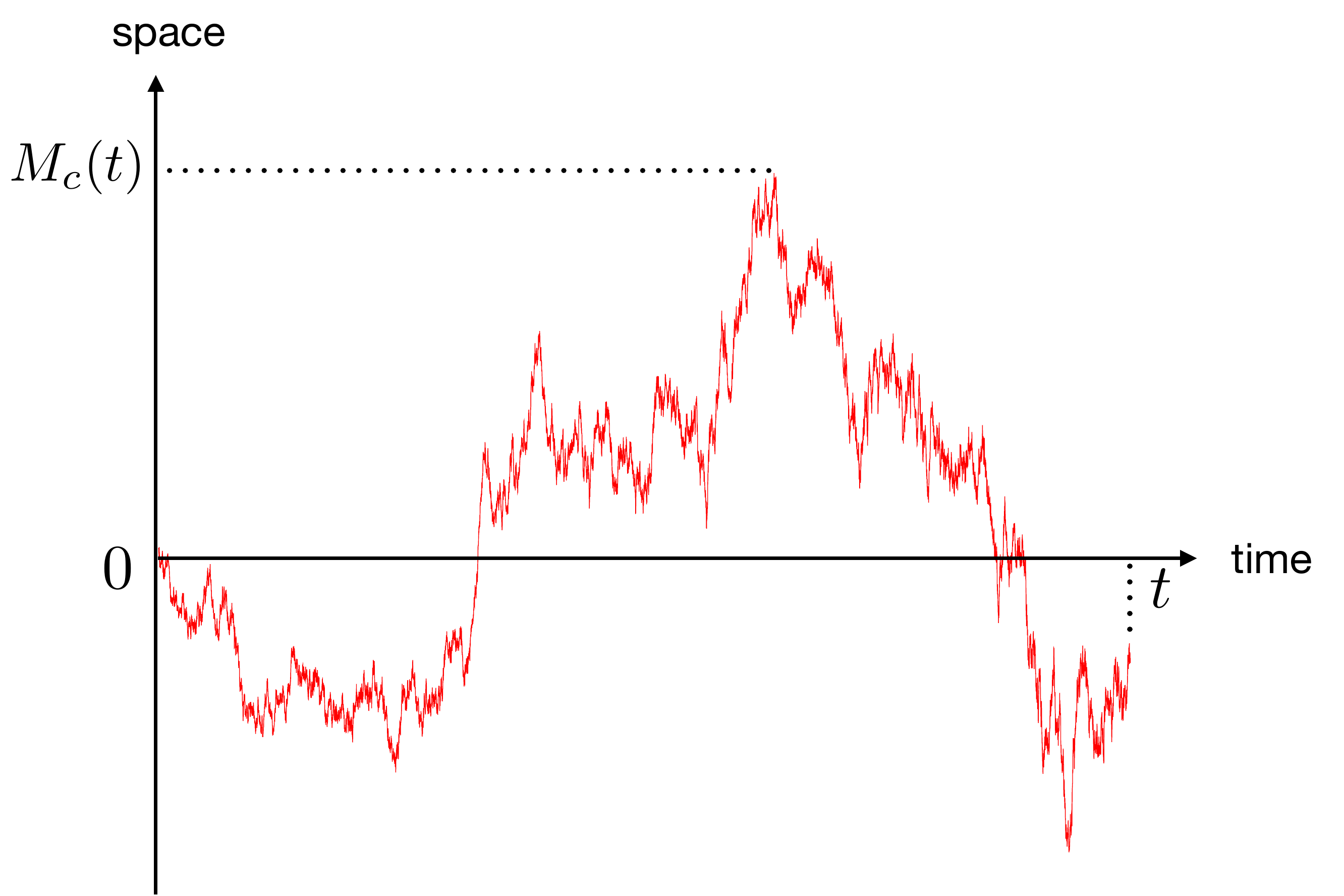}
\caption{Trajectory of a Brownian motion with a drift $c$, together with its maximum $M_c(t)$, over the interval $[0,t]$.}\label{fig_bm}
\end{figure}

In this section, we derive the exact expression\ (\ref{scaling_mu2_1}) for the expected maximum of a $1D$ Brownian motion in the presence of a constant drift. We consider a biased Brownian motion on a line, $x(t)$, starting from $x=0$ at $t=0$ and evolving as
\begin{equation}
\frac{dx(t)}{dt}= c + \sqrt{2D}\, \eta(t)
\label{lange.1}
\end{equation}
where $c$ is the drift, $D$ is the diffusion constant, and $\eta(t)$ is a Gaussian white noise with zero mean and delta correlator, $\langle \eta(t)\eta(t')\rangle = \delta(t-t')$. Let $M_c(t)$ denote the maximum of this process up to time $t$ (see also Fig. \ref{fig_bm})
\begin{equation}
M_c(t) = \max_{0\le \tau\le t}\{x(\tau)\} \,.
\label{max.1}
\end{equation}
First, we compute the cumulative distribution of $M_c(t)$, denoted by ${\rm Prob}[M_c(t)\le z]\equiv Q_c(z,t)$, where the subscript $c$ denotes the presence of the constant drift $c$. Clearly, $Q_c(z,t)$ is also the probability that the Brownian trajectory stays below $z\ge 0$ up to time $t$. To compute this, it is convenient to define a new random variable, $y(t)= z-x(t)$, so that $y(t)$ represents a Brownian motion with a drift $-c$ starting from $y=z$ at $t=0$. Hence, $Q_c(z,t)$ is the probability that the process $y(t)$ (with drift $-c$) stays positive (does not cross zero) up to time $t$. It is then easy to write a backward Fokker-Planck evolution for $Q_c(z,t)$~\cite{pers_review},
\begin{equation}
\frac{\partial Q_c(z,t)}{\partial t}= D\, \frac{\partial^2 Q_c}{\partial 
z^2} - c\, \frac{\partial Q_c}{\partial z}  \, ,
\label{bfp_max.1}
\end{equation}
valid for $z\ge 0$ with the boundary conditions
\begin{equation}
Q_c(z=0,\, t)=0\, ; \quad {\rm and} \quad Q_c(z\to \infty,\, t)=1
\label{bc.1}
\end{equation}
and the initial condition
\begin{equation}
Q_c(z,\,  t=0)=1 \quad {\rm for}\,\, z>0 \, .
\label{ic.1}
\end{equation}
This linear equation can be solved exactly by first taking Laplace transform with respect to $t$, solving the resulting ordinary differential equation and Laplace inverting back to $t$. Another derivation, using a slightly different method, was given in Ref.~\cite{MC02}. Both methods yield the result, valid for any $t$,
\begin{equation}
Q_c(z,t)= \frac{1}{2}\left[{\rm erfc}\left(- 
\frac{z-ct}{\sqrt{4Dt}}\right)- 
e^{c\,z/D}\, {\rm erfc}\left(\frac{z+ct}{\sqrt{4Dt}}\right)\right]\, ,
\label{sol_max.1}
\end{equation}
where ${\rm erfc}(x)= \frac{2}{\sqrt{\pi}}\, \int_x^{+\infty} e^{-u^2}\, du$ is the complementary error function. It is easy to see from Eq. (\ref{sol_max.1}) that for $c>0$, the cumulative distribution $Q_{c>0}(z,t)$ is always time-dependent, while for $c<0$, it approaches a time-independent stationary distribution
\begin{equation}
Q_{c<0}(z,t) \xrightarrow[t\to \infty]{} 1- \exp\left[- \frac{|c|\, 
z}{D}\right]\,.
\label{stat_dist.1}
\end{equation}
The PDF of $M_c(t)$ is the derivative $\partial_z Q_c(z,t)$, hence the expected maximum is given by $\E[M_c(t)]= \int_0^{+\infty} z\, \partial_z Q_c(z,t)\, dz$, which can be expressed, via integration by parts, as
\begin{equation}
\E[M_c(t)] = \int_0^{+\infty} \left[1- Q_c(z,t)\right]\, dz \,.
\label{mean_max.1}
\end{equation}
From Eq.\ (\ref{mean_max.1}) and the result in Eq. (\ref{sol_max.1}), it is possible to compute the expected maximum $\E[M_c(t)]$ exactly at any time $t$. First, we write Eq. (\ref{bfp_max.1}) as
\begin{equation}\label{mean_max_new.2}
\frac{\partial Q_c(z,t)}{\partial t}=- \frac{\partial J_c(z,t)}{\partial z},
\end{equation}
where
\begin{equation}\label{mean_max_new.2bis}
J_c(z,t)=-D\, 
\frac{\partial Q_c}{\partial z} + c\, Q_c(z,t)\, .
\end{equation}
Using the explicit solution in Eq. (\ref{sol_max.1}) on the right-hand side of\ (\ref{mean_max_new.2bis}), it is easily checked that
\begin{equation}
J_c(z,t)= - \sqrt{\frac{D}{\pi t}}\, \exp\left[-\frac{(ct-z)^2}{4Dt}\right] + \frac{c}{2}\, 
{\rm erfc}\left(\frac{ct-z}{\sqrt{4Dt}}\right)\, .
\label{current.1}
\end{equation}
Then, we take the time derivative of Eq. (\ref{mean_max.1}) in which we use the relation in Eq. (\ref{mean_max_new.2}). This  gives
\begin{equation}
\frac{d \, \E[M_c(t)]}{dt}= \int_0^{+\infty} dz\, \left[\frac{\partial J_c(z,t)}{\partial z}\right]= J_c(+\infty,t)-J_c(0,t)\, .
\label{mean_max_new.3}
\end{equation}
From Eq. (\ref{current.1}) it follows that $J_c(+\infty, t)= c$ and
\begin{equation*}
J_c(0,t)= \frac{c}{2}\, {\rm erfc}\left(\frac{c\sqrt{t}}{\sqrt{4D}}\right)
-\sqrt{\frac{D}{\pi t}}\, e^{-c^2 t/{4D}}\, ,
\end{equation*}
which, together with Eq. (\ref{mean_max_new.3}), yields
\begin{equation}
\frac{d \,\E[M_c(t)]}{dt}= c - \frac{c}{2}\,
{\rm erfc}\left(\frac{c\sqrt{t}}{\sqrt{4D}}\right) + \sqrt{\frac{D}{\pi t}}\, e^{-c^2 t/{4D}}\, .
\label{mean_max_new.4}
\end{equation}
Now, by the identity ${\rm erfc}(x)+ {\rm erfc}(-x)=2$ (for all $x$) one has
\begin{equation*}
c- \frac{c}{2}\,
{\rm erfc}\left(\frac{c\sqrt{t}}{\sqrt{4D}}\right)= c\, \theta(c) - \frac{|c|}{2}\,
{\rm erfc}\left(\frac{|c|\sqrt{t}}{\sqrt{4D}}\right)\, ,
\end{equation*}
where $\theta(c)$ is the Heaviside step function [$\theta(c)=1$ for $c>0$ and $\theta(c)=0$ for $c<0$], and Eq.\ (\ref{mean_max_new.4}) can be rewritten as
\begin{equation}
\frac{d \E[M_c(t)]}{dt}= c\, \theta(c) - \frac{|c|}{2}\,
{\rm erfc}\left(\frac{|c|\sqrt{t}}{\sqrt{4D}}\right) +\sqrt{\frac{D}{\pi t}}\, e^{-c^2 t/{4D}}\, .
\label{mean_max.5}
\end{equation}
Note that, for $c=0$, one gets, by integrating over $t$, the result just stated in Eq. (\ref{BM_c=0}).  
Unlike the expression in Eq.\ (\ref{mean_max_new.4}), only the first -- pure drift -- term on the right-hand side of Eq.\ (\ref{mean_max.5}) depends on the sign of $c$. Finally, integrating\ (\ref{mean_max.5}) with respect to $t$ and using $\E[M_c(0)]=0$, we get an exact expression for $\E[M_c(t)]$, valid for all $c$ and all $t$,
\begin{equation}
\E[M_c(t)] = c\, \theta(c)\, t + \sqrt{D\, t}\, {\cal G}_2\left(\frac{|c|\sqrt{t}}{\sqrt{D}}\right),
\label{mean_max_scaling.1}
\end{equation}
where the scaling function ${\cal G}_2(u)$ is exactly given by
\begin{eqnarray}\label{mean_max_scaling.2}
{\cal G}_2(u)&=&
\frac{4}{u\sqrt{\pi}}\, \int_0^{u/2} dv\, \left[e^{-v^2}- \sqrt{\pi}\, v\, {\rm erfc}(v)\right] \nonumber \\
&=&\frac{1}{\sqrt{\pi}}\, e^{-u^2/4} + \frac{1}{u}\, {\rm erf}\left(\frac{u}{2}\right)- 
\frac{u}{2}\, {\rm erfc}\left(\frac{u}{2}\right)\, ,
\end{eqnarray}
with the asymptotics ${\cal G}_2(u) \to 2/\sqrt{\pi}$ as $u\to 0$ and ${\cal G}_2(u) \to 1/u$ as $u \to \infty$. In particular, at late times
$t\to \infty$, it follows that
\begin{equation}
\E[M_{c \neq 0}(t)] \sim \theta(c)\, c\, t + \frac{D}{|c|}\ \ \ \ \ (t \to \infty) \;,
\label{mean_max.2}
\end{equation}
as stated in Eq.\ (\ref{BM_cneq0}). Thus, for a positive drift $c>0$, the expected maximum increases linearly with $t$ (with speed $c$), while for a negative drift $c<0$, it approaches a constant $D/|c|$ as $t\to \infty$. The equations.\ (\ref{mean_max_scaling.1}) and\ (\ref{mean_max_scaling.2}) coincide respectively with Eqs.\ (\ref{scaling_mu2_1}) and\ (\ref{scaling_mu2}) in the introduction.
%
%
\section{Expected maximum of $\bm{1D}$ discrete time random walks with a constant drift: the analytical tools}\label{sec2}
In this section, we develop the analytical tools needed to derive the asymptotic behaviors of the expected maximum for a discrete time random walk with a non-zero drift as defined in Eq.\ (\ref{eq1.1}). To begin with, we generalize the so-called Hopf-Ivanov formula\ \cite{Ivanov} to the case of a biased random walks. We follow the same line as in our derivation of the formula for unbiased walk in the Appendix A of Ref. \cite{MMS2014}. The results in Secs.\ \ref{sec2.1} and\ \ref{sec2.2}, as well as the generalized Pollaczek-Spitzer formula (\ref{eq3.5}), hold for $0<\mu\le 2$.
%
%
\subsection{The Hopf-Ivanov formula in the presence of a drift}\label{sec2.1}

\begin{figure}
\includegraphics[width = 0.7 \linewidth]{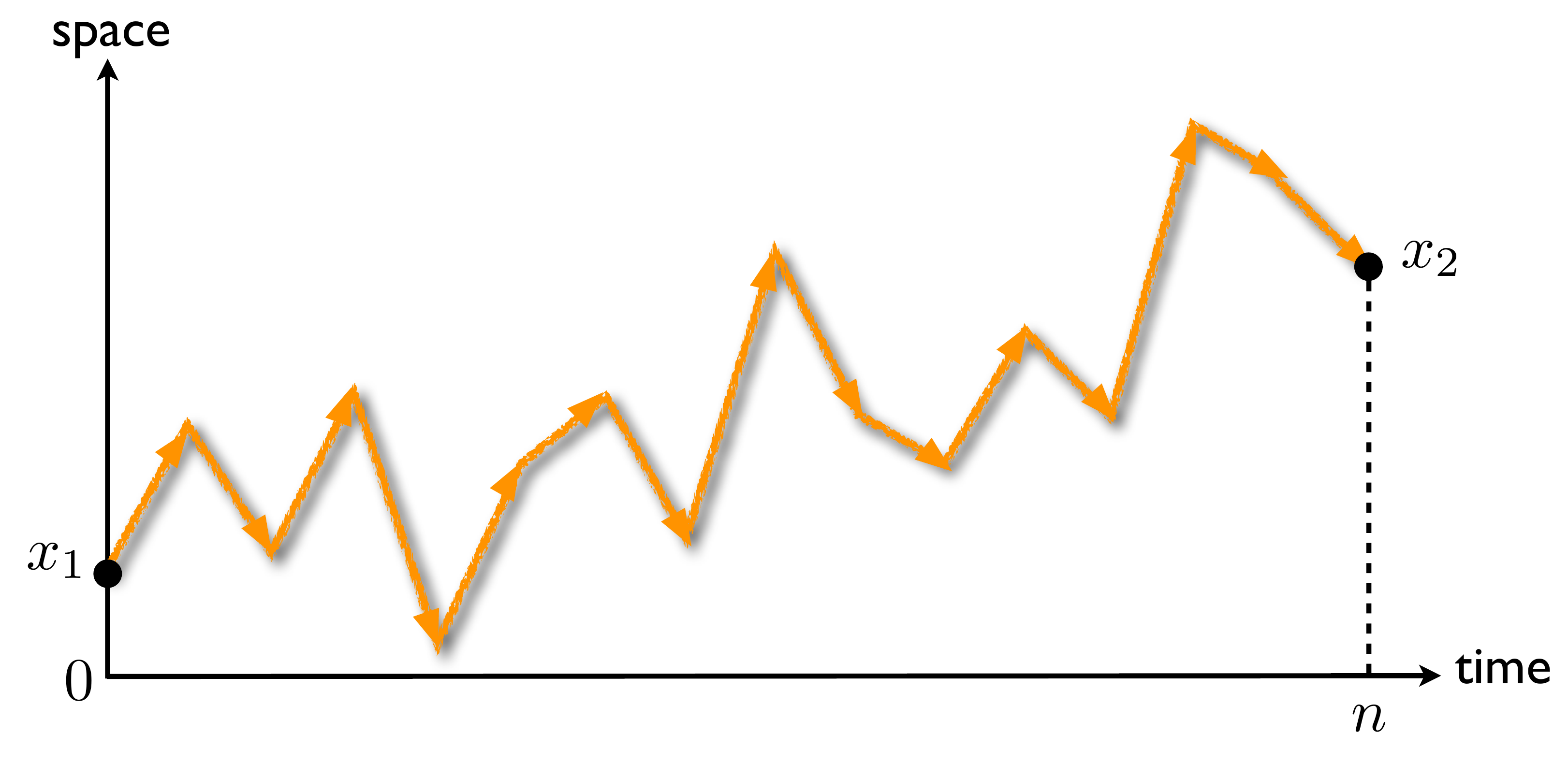}
\caption{Trajectory of a random walk contributing to the constrained propagator $p^{(c)}_{>0}(x_2,n\vert x_1,0)$ for the walker starting at $x_1\ge 0$, arriving at $x_2\ge 0$ after $n$ steps, and always staying on the positive side in between.}\label{fig_propag}
\end{figure}

Consider the biased random walk defined in Eq.\ (\ref{eq1.1}). Let $p^{(c)}_{>0}(x_2,n\vert x_1,0)$ denote the Green's function (propagator) for the walker starting at $x_1\ge 0$, arriving at $x_2\ge 0$ after $n$ steps, and always staying on the positive side in between (Fig. \ref{fig_propag}). Assume that the walker reaches the minimum $y\ge 0$ of the walk at time $n_1$ and split the walk into two successive parts: from $x_1$ at time $0$ to $y$ at time $n_1$, then from $y$ at time $n_1$ to $x_2$ at time $n$. By taking $y$ as a new origin in both parts, reversing time direction in the first one, and taking into account all the possible values of $y$ and $n_1$, one has
\begin{eqnarray}\label{eq2.1.1}
&&p^{(c)}_{>0}(x_2,n\vert x_1,0)=\sum_{n_1=1-\sigma}^{n-\sigma}\int_0^{\min(x_1,x_2)}
p^{(-c)}_{>0}(x_1-y,n_1\vert 0,0)p^{(c)}_{>0}(x_2-y,n-n_1\vert 0,0)\, dy, \nonumber \\
&&=\sum_{n_1\ge 1-\sigma}\sum_{n_2\ge\sigma}\left\lbrack\int_0^{\min(x_1,x_2)}
p^{(-c)}_{>0}(x_1-y,n_1\vert 0,0)p^{(c)}_{>0}(x_2-y,n_2\vert 0,0)\, dy\right\rbrack\, \delta_{n_1+n_2,n},
\end{eqnarray}
where $n_2=n-n_1$ and $\sigma =1$ (resp. $0$) if $\min(x_1,x_2)=x_1$ (reps. $x_2$). Note that reversing time direction changes the sign of the drift (i.e., turns $c$ into $-c$). The sums on the right-hand side of\ (\ref{eq2.1.1}) can actually be continued to $n_1,n_2\ge 0$, (the difference corresponds to a term proportional to $\delta\lbrack y-\max(x_1,x_2)\rbrack$ in the integral over $y$ the contribution of which is zero), and\ (\ref{eq2.1.1}) reads
\begin{equation}\label{eq2.1.2}
p^{(c)}_{>0}(x_2,n\vert x_1,0)=
\sum_{n_1\ge 0}\sum_{n_2\ge 0}\left\lbrack\int_0^{\min(x_1,x_2)}
p^{(-c)}_{>0}(x_1-y,n_1\vert 0,0)p^{(c)}_{>0}(x_2-y,n_2\vert 0,0)\, dy\right\rbrack\, \delta_{n_1+n_2,n}.
\end{equation}
Let
\begin{equation}\label{eq2.1.3}
G^{(c)}_{>0}(x_2,x_1,s)=\sum_{n\ge 0}p^{(c)}_{>0}(x_2,n\vert x_1,0)s^n.
\end{equation}
From\ (\ref{eq2.1.2}) and\ (\ref{eq2.1.3}) one gets
\begin{eqnarray}\label{eq2.1.4}
G^{(c)}_{>0}(x_2,x_1,s)&=&\int_0^{\min(x_1,x_2)}
\left(\sum_{n_1\ge 0}p^{(-c)}_{>0}(x_1-y,n_1\vert 0,0)s^{n_1}\right)
\left(\sum_{n_2\ge 0}p^{(c)}_{>0}(x_2-y,n_2\vert 0,0)s^{n_2}\right)\, dy \nonumber \\
&=&\int_0^{\min(x_1,x_2)}G^{(-c)}_{>0}(x_1-y,0,s)G^{(c)}_{>0}(x_2-y,0,s)\, dy.
\end{eqnarray}
Laplace transforming\ (\ref{eq2.1.4}) with respect to both $x_1$ and $x_2$, and using\ (\ref{eq2.1.3}), one gets
\begin{equation}\label{eq2.1.5}
\int_0^{+\infty}\int_0^{+\infty}
\sum_{n\ge 0}p_{>0}^{(c)}(x_2,n\vert x_1,0)s^n\, 
{\rm e}^{-\lambda_1 x_1-\lambda_2 x_2}dx_1\, dx_2=
\frac{\phi^{(-c)}(\lambda_1 ,s)\phi^{(c)}(\lambda_2 ,s)}{\lambda_1+\lambda_2},
\end{equation}
where $\phi^{(c)}(\lambda ,s)$ is the Laplace transform of $G^{(c)}_{>0}(x,0,s)$ with respect to $x$. Its expression is derived in Appendix\ \ref{app1}. One finds
\begin{equation}\label{eq2.1.6}
\phi^{(c)}(\lambda ,s)=\exp\left(-\frac{1}{2\pi}\int_{-\infty}^{+\infty}
\frac{\ln\lbrack 1-s\hat{f}(k){\rm e}^{ikc}\rbrack}{\lambda +ik}\, dk\right).
\end{equation}
Equations\ (\ref{eq2.1.5}) and\ (\ref{eq2.1.6}) generalize the Hopf-Ivanov formula to random walks with a drift as defined in Eq.\ (\ref{eq1.1}).
%
%
\subsection{An alternative expression for $\bm{\phi^{(c)}(\lambda ,s)}$}\label{sec2.2}
In addition to the integral representation\ (\ref{eq2.1.6}) of $\phi^{(c)}(\lambda ,s)$, there is an alternative formulation in which $\phi^{(c)}(\lambda ,s)$ can be written as an exponential of an auxiliary generating function written as a power series of $s$. This new, power series, representation is obtained by expanding the logarithm on the right-hand side of\ (\ref{eq2.1.6}) as a power series of $s$ and by using the Laplace transform
\begin{equation}\label{eq2.2.1}
\int_0^{+\infty}g(x-y){\rm e}^{-px}\, dx =
\frac{1}{2\pi}\int_{-\infty}^{+\infty}\frac{\hat{g}(k)\exp(iky)}{p+ik}\, dk.
\end{equation}
One gets
\begin{eqnarray}\label{eq2.2.2a}
\ln\phi^{(c)}(\lambda ,s)&=&-\frac{1}{2\pi}\int_{-\infty}^{+\infty}
\frac{\ln\lbrack 1-s\hat{f}(k){\rm e}^{ikc}\rbrack}{\lambda +ik}\, dk \nonumber \\
&=&\frac{1}{2\pi}\sum_{m=1}^{+\infty}\frac{s^m}{m}\int_{-\infty}^{+\infty}
\frac{\hat{f}(k)^m{\rm e}^{inkc}}{\lambda +ik}\, dk \nonumber \\
&=&\sum_{m=1}^{+\infty}\frac{s^m}{m}\int_0^{+\infty}
{\cal F}_{\mu ,m}\left(x-m^{1-1/\mu}c\right){\rm e}^{-m^{1/\mu}\lambda x}\, dx ,
\end{eqnarray}
where we have made the change of variable $x\rightarrow xm^{1/\mu}$ in the series, and
\begin{equation}\label{eq2.2.3a}
{\cal F}_{\mu ,m}(x)=\int_{-\infty}^{+\infty}\hat{f}\left(\frac{q}{m^{1/\mu}}\right)^m
{\rm e}^{-iqx}\, \frac{dq}{2\pi},
\end{equation}
is the PDF of the sum $m^{-1/\mu}\sum_{i=1}^m \eta_i$. Note in particular that in the limit $m \to \infty$, ${\cal F}_{\mu ,m}(x)$ converges towards to a stable distribution of index $\mu$, i.e.
\begin{eqnarray}\label{Fmu_asympt}
\lim_{m \to \infty} {\cal F}_{\mu ,m}(x) = {\cal F}_{\mu ,\infty}(x) = \frac{1}{a} \int_{-\infty}^{+\infty} \frac{dq}{2\pi} e^{-|q|^\mu - i q x/a} = \frac{1}{a} f_{S,\mu}\left( \frac{x}{a}\right) \;,
\end{eqnarray}
where we recall that $a$ is such that $\hat f(q) \approx 1 - |a\,q|^\mu$ for small $q$. Writing then the equation\ (\ref{eq2.2.2a}) as
\begin{equation}\label{eq2.2.2b}
\phi^{(c)}(\lambda ,s)=\exp\left(\sum_{m=1}^{+\infty}\alpha_{\mu ,m}(c,\lambda)\, s^m\right),
\end{equation}
with
\begin{equation}\label{eq2.2.3b}
\alpha_{\mu ,m}(c,\lambda)=\frac{1}{m}\int_0^{+\infty}
{\cal F}_{\mu ,m}\left(x-m^{1-1/\mu}c\right){\rm e}^{-m^{1/\mu}\lambda x}\, dx,
\end{equation}
one obtains an alternative expression for $\phi^{(c)}(\lambda ,s)$ equivalent to the one in Eq.\ (\ref{eq2.1.6}) in which the integral over $k$ is replaced with the generating function of the sequence $\alpha_{\mu ,m}(c,\lambda)$. 

This alternative expression for $\phi^{(c)}(\lambda ,s)$ in Eqs. (\ref{eq2.2.2b}) and (\ref{eq2.2.3b}) will prove useful later in section \ref{sec4} for the derivation of the asymptotics of the expected maximum in the presence of a drift. As another useful application of this formalism, let us point out that, using the generalized Hopf-Ivanov formula (\ref{eq2.1.5}) and the representation of $\phi^{(c)}(\lambda ,s)$ in Eqs. (\ref{eq2.2.2b}) and (\ref{eq2.2.3b}), one can derive the generalized Sparre Andersen theorem for the survival probability of the walk in the presence of a drift\ \cite{SA1954}. We provide this derivation in Appendix \ref{App_SA}.  
%
%
\subsection{Generalized Pollaczek-Spitzer formula and generating function of $\bm{\E[M_c(n)]}$}\label{sec3}
In this section we derive the generating function of the expected maximum, $\E[M_c(n)]$, from the generalization of the Pollaczek-Spitzer formula to random walks with a drift.

Let
\begin{equation}\label{eq3.1}
p_c(\m,n)=\frac{d\,{\rm Prob}(M_c(n)<\m)}{d\m},
\end{equation}
denote the PDF of $M_c(n)$ and write
\begin{equation}\label{eq3.2}
\E\left({\rm e}^{-\lambda M_c(n)}\right)=\int_0^{+\infty}p_c(\m,n){\rm e}^{-\lambda \m}d\m,
\end{equation}
its Laplace transform. Substituting (\ref{eq3.1}) in (\ref{eq3.2}), doing integration by parts, and using ${\rm Prob}(M_c(n)<0)=0$ gives
\begin{eqnarray}\label{eq3.2-inter}
\E\left({\rm e}^{-\lambda M_c(n)}\right) = \lambda \int_0^{+\infty} \e^{-\lambda \m} \, {\rm Prob}(M_c(n)<\m) \, d\m  \;.
\end{eqnarray}
Let $p_{<\m}^{(c)}(x,n\vert 0,0)$ denote the probability density that the walker arrives at $x$ after $n$ steps, starting from $0$ at step $0$ and always staying below the level $\m$ in between. If we integrate over the arrival point $x$ at step $n$ for $x \in (-\infty,\m]$ this just gives the probability that the maximum $M_c(n)$ of the trajectory is less than $\m$. Hence
\begin{eqnarray}\label{eq3.3}
{\rm Prob}(M_c(n)<\m)&=&\int_{-\infty}^{\m} p_{<\m}^{(c)}(x,n\vert 0,0)\, dx \nonumber \\
&=&\int_0^{+\infty} p_{>0}^{(-c)}(x,n\vert \m,0)\, dx \;,
\end{eqnarray}
where, in the second line, we have performed the change of variable $x \to \m -x$. This automatically
changes the sign of the drift, as follows from Eq.~(\ref{eq1.1}). Using then\ (\ref{eq3.3}) on the right-hand side of\ (\ref{eq3.2-inter}), one gets 
\begin{equation}\label{eq3.4}
\sum_{n\ge 0}\E\left({\rm e}^{-\lambda M_c(n)}\right)s^n=
\lambda\int_0^{+\infty}\int_0^{+\infty}\sum_{n\ge 0}p_{>0}^{(-c)}(x,n\vert \m,0)s^n \,
{\rm e}^{-\lambda \m}dx\, d\m,
\end{equation}
which reads, according to the generalized Hopf-Ivanov formula\ (\ref{eq2.1.5}) with $\lambda_1 =\lambda$ and $\lambda_2 =0$,
\begin{equation}\label{eq3.5}
\sum_{n\ge 0}\E\left({\rm e}^{-\lambda M_c(n)}\right)s^n
=\phi^{(c)}(\lambda ,s)\phi^{(-c)}(0,s).
\end{equation}
Equation\ (\ref{eq3.5}) generalizes the Pollaczek-Spitzer formula to random walks with a drift (see Eq. (34) of Ref. \cite{AS2005} and references therein for the case without drift). It is shown in Appendix \ref{App_Spitzer} that our formula (\ref{eq3.5}) is equivalent to the one given by Spitzer in Eq. (3.1) of Ref. \cite{Spi57}.  

The generating function for the expected maximum is then readily obtained by differentiation of\ (\ref{eq3.5}) with respect to $\lambda$. One has
\begin{eqnarray}\label{eq3.6a}
\sum_{n\ge 0}\E[M_c(n)]s^n
&=&-\sum_{n\ge 0}\left.\frac{\partial}{\partial\lambda}\E\left({\rm e}^{-\lambda M_c(n)}\right)
\right\vert_{\lambda =0}s^n
\nonumber \\
&=&-\phi^{(-c)}(0,s)
\left.\frac{\partial\phi^{(c)}(\lambda ,s)}{\partial\lambda}\right\vert_{\lambda =0} \nonumber \\
&=&-\phi^{(-c)}(0,s)\phi^{(c)}(0,s)
\left.\frac{\partial\ln\phi^{(c)}(\lambda ,s)}{\partial\lambda}\right\vert_{\lambda =0},
\end{eqnarray}
and by using the expression\ (\ref{eq2.1.6}) of $\phi^{(c)}(\lambda ,s)$ on the right-hand side of\ (\ref{eq3.6a}), one finally gets
\begin{equation}\label{eq3.6b}
\sum_{n\ge 0}\E[M_c(n)]s^n =\frac{\Psi^{(c)}(s)}{1-s},
\end{equation}
where we have used $\phi^{(-c)}(0,s)\phi^{(c)}(0,s) =\lim_{\lambda\to 0}\phi^{(-c)}(\lambda ,s)\phi^{(c)}(\lambda ,s)=(1-s)^{-1}$ and
\begin{equation}\label{eq3.7}
\Psi^{(c)}(s)=\frac{1}{2\pi}\frac{\partial}{\partial\lambda}\left.\int_{-\infty}^{+\infty}
\frac{\ln\lbrack 1-s\hat{f}(k){\rm e}^{ikc}\rbrack}{\lambda +ik}\, dk\right\vert_{\lambda =0},
\end{equation}
where $\lambda$ goes to zero from ${\rm Re}(\lambda)=+\infty$. An alternative expression for $\Psi^{(c)}(s)$, equivalent to the one in Eq.\ (\ref{eq3.7}), can be obtained from the power series representation of $\phi^{(c)}(\lambda ,s)$ in Eqs.\ (\ref{eq2.2.2b})-(\ref{eq2.2.3b}). After some straightforward algebra, one finds
\begin{eqnarray}\label{eq3.8}
\Psi^{(c)}(s)&=&-\sum_{m=1}^{+\infty}s^m\left.\frac{\partial\alpha_{\mu ,m}(c,\lambda)}{\partial\lambda}
\right\vert_{\lambda =0} \nonumber \\
&=&\sum_{m=1}^{+\infty}\frac{s^m}{m^{1-1/\mu}}\int_0^{+\infty}
x \,{\mathcal F}_{\mu ,m}\left(x-m^{1-1/\mu}c\right)\, dx \;,
\end{eqnarray}
where ${\mathcal F}_{\mu,m}(x)$ is given in Eq. (\ref{eq2.2.3a}). 

In principle, the equations\ (\ref{eq3.6b}) and\ (\ref{eq3.7}) [or\ (\ref{eq3.8})] solve the problem for any value of $n$. In particular, identifying the powers of $s$ on both sides of Eq. (\ref{eq3.6b}) in which $\Psi^{(c)}(s)$ is written as\ (\ref{eq3.8}), one obtains an exact expression for $\E[M_c(n)]$, valid for any $n \geq 1$,
\begin{eqnarray}\label{exact_E}
\E[M_c(n)] = \sum_{m=1}^n m^{1/\mu-1} \int_0^{+\infty}
x \,{\mathcal F}_{\mu ,m}\left(x-m^{1-1/\mu}c\right)\, dx \;,
\end{eqnarray}
which will be useful in the following for comparison with numerical data. Note  that, taking advantage of all the ${\mathcal F}_{\mu ,m}$s being symmetric and normalized to unity, the relation in Eq. (\ref{exact_E}) can also be written as
\begin{eqnarray}\label{exact_E_2}
\E[M_c(n)] = c \, \theta(c)\, n + \sum_{m=1}^n m^{1/\mu-1} \int_0^{+\infty} x {\mathcal F}_{\mu ,m}\left(x + m^{1-1/\mu}|c|\right)\, dx \;.
\end{eqnarray}
The first term on the right-hand side of\ (\ref{exact_E_2}) corresponds to the pure drift behavior which is thus made explicit by this formulation.The departure from the pure drift behavior given by the second term on the right-hand side of\ (\ref{exact_E_2}) does not depend on the sign of $c$, in agreement with Eq.\ (\ref{identity_av}). Finally, by using the Fourier representation\ (\ref{eq2.2.3a}) of ${\mathcal F}_{\mu ,m}(x)$ on the right-hand side of Eq.\ (\ref{exact_E_2}) and after some straightforward algebra, one obtains the equivalent alternative expression
\begin{eqnarray}\label{exact_E_3}
\E[M_c(n)] = c \, \theta(c)\, n  + \sum_{m=1}^n \int_{|c|\,m}^{+\infty} dy \left(\frac{y}{m}-|c| \right) \int_{-\infty}^{+\infty} \left[\hat f(q)\right]^m \e^{-iq\,y} \frac{dq}{2\pi} \;.
\end{eqnarray}

Besides these exact expressions, we will see below that the large $n$ behavior of $\E[M_c(n)]$ can be conveniently obtained by analyzing the right-hand side of Eq.~(\ref{eq3.6b}) near its dominant singularity by means of appropriate Tauberian theorems. However, extracting this asymptotic behavior is still nontrivial and this is the subject of the next two sections.
%
%
\section{Large $\bm{n}$ behavior of $\bm{\E[M(n)]}$ for symmetric discrete time random walks ($\bm{c=0}$)}\label{sec4}
In the absence of a drift ($c=0$), the large $n$ behavior of $\E[M(n)]$ has already been extensively studied by Comtet and Majumdar in Ref. \cite{AS2005}, starting from the same $k$-integral representation as in our Eq.\ (\ref{eq3.7}). In this section, we show how their results can also be obtained from the alternative, power series, representation in Eq.\ (\ref{eq3.8}). For simplicity, in the rest of this section we will restrict ourselves to the case $\nu >\mu +1$ [see Eq. (\ref{eq1.3})], though our results can be generalized also to $\mu<\nu<\mu +1$.

By the central limit theorem for the sum $m^{-1/\mu}\sum_{i=1}^{m}\eta_i$ one has ${\cal F}_{\mu ,m}(x)\sim a^{-1}f_{S,\mu}(x/a)$ as $m\to +\infty$, where $f_{S,\mu}(x)$ is the stable law with Fourier transform $\hat{f}_{S,\mu}(k)=\exp(-\vert k\vert^\mu)$, and it is useful to extract from $\Psi^{(0)}(s)$ the contribution of the asymptotic stable distribution. This is done by adding and subtracting in\ (\ref{eq3.7}) or\ (\ref{eq3.8}) with $c=0$ the expression of $\Psi^{(0)}(s)$ in which ${\cal F}_{\mu ,m}(x)$ is replaced with $a^{-1}f_{S,\mu}(x/a)$. One obtains
\begin{equation}\label{eq3.9}
\Psi^{(0)}(s)=\Psi_1^{(0)}(s)+\Psi_2^{(0)}(s),
\end{equation}
with, using Eq.\ (\ref{eq3.7}),
\begin{equation}\label{eq3.10a}
\Psi_1^{(0)}(s)=\frac{1}{2\pi}\int_{-\infty}^{+\infty}\ln\left(
\frac{1-s\hat{f}(k)}{1-s{\rm e}^{-\vert ak\vert^\mu}}\right)
\, \frac{dk}{k^2},
\end{equation}
and
\begin{equation}\label{eq3.10b}
\Psi_2^{(0)}(s)=\frac{1}{2\pi}\frac{\partial}{\partial\lambda}\left.\int_{-\infty}^{+\infty}
\frac{\ln(1-s{\rm e}^{-\vert ak\vert^\mu})}{\lambda +ik}\, dk\right\vert_{\lambda =0},
\end{equation}
where $\lambda$ goes to zero from ${\rm Re}(\lambda)=+\infty$, or, equivalently, using Eq.\ (\ref{eq3.8}) instead of Eq.\ (\ref{eq3.7}),
\begin{equation}\label{eq3.11a}
\Psi_1^{(0)}(s)=\sum_{m=1}^{+\infty}\frac{s^m}{m^{1-\frac{1}{\mu}}}\int_0^{+\infty}
x\left\lbrack{\mathcal F}_{\mu ,m}(x)-\frac{1}{a}f_{S,\mu}\left(\frac{x}{a}\right)
\right\rbrack\, dx,
\end{equation}
and
\begin{equation}\label{eq3.11b}
\Psi_2^{(0)}(s)=\sum_{m=1}^{+\infty}\frac{s^m}{m^{1-1/\mu}}\int_0^{+\infty}
\frac{x}{a}f_{S,\mu}\left(\frac{x}{a}\right)\, dx.
\end{equation}
It is clear that $\Psi_1^{(0)}(s)$ is entirely due to the departure of the jump distribution from the stable law with same $\mu$ [i.e., $\Psi_1^{(0)}(s)=0$ for $f(x)=a^{-1}f_{S,\mu}(x/a)$]. From the equation (\ref{eq3.10a}) and the small $k$ behavior in Eq.\ (\ref{eq1.3}) with $\nu>\mu +1$, it is also clear that $\Psi_1^{(0)}(1)$ exists with
\begin{equation}\label{eq3.12}
\Psi_1^{(0)}(s)=\Psi_1^{(0)}(1)+O(1-s)\ \ \ \ \ (s\rightarrow 1).
\end{equation}
From\ (\ref{eq3.9}) and\ (\ref{eq3.11b}), one has
\begin{eqnarray}\label{eq4.1}
\Psi^{(0)}(s)&=&\Psi_1^{(0)}(s)
+a \,{\rm Li}_{1-1/\mu}(s)\int_0^{+\infty}yf_{S,\mu}(y)\, dy \nonumber \\
&=&\Psi_1^{(0)}(s)
+\frac{a\,\Gamma(1-1/\mu)}{\pi}{\rm Li}_{1-1/\mu}(s),
\end{eqnarray}
where ${\rm Li}_\alpha(z)=\sum_{m=1}^{+\infty}m^{-\alpha}z^m$. The asymptotic behavior of\ (\ref{eq4.1}) near the dominant singularity at $s=1$ is determined by the one of ${\rm Li}_{1-1/\mu}(s)$, [according to\ (\ref{eq3.12}), the contribution of $\Psi_1^{(0)}(s)$ reduces to the constant $\Psi_1^{(0)}(1)$], and from
\begin{equation}\label{eq4.2}
{\rm Li}_{1-1/\mu}(s)=\zeta(1-1/\mu)+\frac{\Gamma(1/\mu)}{(1-s)^{1/\mu}}
+O((1-s)^{1-1/\mu})\ \ \ \ \ (s\rightarrow 1),
\end{equation}
where $\zeta(z)$ is the Riemann zeta function, one finds
\begin{equation}\label{eq4.3}
\Psi^{(0)}(s)= a\gamma +\frac{a\Gamma(1-1/\mu)\Gamma(1/\mu)}{\pi (1-s)^{1/\mu}}
+O((1-s)^{1-1/\mu})\ \ \ \ \ (s\rightarrow 1),
\end{equation}
with
\begin{equation}\label{eq4.4}
\gamma=\frac{\Psi_1^{(0)}(1)}{a}+\frac{\Gamma(1-1/\mu)}{\pi}\zeta(1-1/\mu).
\end{equation}
It can be seen immediately from Eq.\ (\ref{eq4.4}) that if the jump distribution $f(\eta)$ is the stable distribution $a^{-1}f_{S,\mu}(\eta/a)$ [i.e. if $\hat{f}(k)=\exp(-\vert ak\vert^\mu)$], then Eq.\ (\ref{eq3.11a}) [or Eq.\ (\ref{eq3.10a})] yields $\Psi_1^{(0)}(1)=0$ and $\gamma$ is exactly given by
\begin{equation}\label{eq4.5}
\gamma =\frac{\Gamma(1-1/\mu)}{\pi}\zeta(1-1/\mu)
=\frac{\zeta(1/\mu)}{(2\pi)^{1/\mu}\sin(\pi/2\mu)},
\end{equation}
where we have used the reflection formulas $\zeta(1-1/\mu)=2(2\pi)^{-1/\mu}\cos(\pi/2\mu)\Gamma(1/\mu)\zeta(1/\mu)$ and $\Gamma(1-1/\mu)\Gamma(1/\mu)=\pi/\sin(\pi/\mu)$. Equation\ (\ref{eq4.5}) coincides with the equation (14) in Ref. \cite{AS2005}. On the other hand, if $f(\eta)$ is not the stable distribution one can use the integral representation\ (\ref{eq3.10a}) for $\Psi_1^{(0)}(1)$ together with the identity
\begin{equation*}
\zeta(1-1/\mu)=\frac{1}{\Gamma(1-1/\mu)}
\int_0^{+\infty}\ln\left(\frac{1-{\rm e}^{-q^\mu}}{q^\mu}\right)\, \frac{dq}{q^2},
\end{equation*}
to write\ (\ref{eq4.4}) in the equivalent integral form
\begin{eqnarray}\label{eq4.6}
\gamma &=&\frac{1}{\pi}\int_{0}^{+\infty}\ln\left(
\frac{1-\hat{f}(q/a)}{1-{\rm e}^{-q^\mu}}\right)
\, \frac{dq}{q^2}
+\frac{\Gamma(1-1/\mu)}{\pi}\zeta(1-1/\mu) \nonumber \\
&=&\frac{1}{\pi}\int_{0}^{+\infty}\ln\left(
\frac{1-\hat{f}(q/a)}{q^\mu}\right)
\, \frac{dq}{q^2},
\end{eqnarray}
which coincides with the equation (13) in Ref. \cite{AS2005}. Injecting\ (\ref{eq4.3}) into\ (\ref{eq3.6b}) with $c=0$, one gets
\begin{equation}\label{eq4.7}
\sum_{n\ge 0}\E[M(n)]s^n =
\frac{a\gamma}{1-s} +\frac{a\Gamma(1-1/\mu)\Gamma(1/\mu)}{\pi (1-s)^{1+1/\mu}}
+O\left(\frac{1}{(1-s)^{1/\mu}}\right)\ \ \ \ \ (s\rightarrow 1),
\end{equation}
and by Darboux's theorem\ \cite{Hen} one finally finds
\begin{equation}\label{eq4.8}
\frac{\E[M(n)]}{a}=\frac{\mu\Gamma(1-1/\mu)}{\pi} n^{1/\mu}+\gamma
+O\left(\frac{1}{n^{1-1/\mu}}\right)\ \ \ \ \ (n\rightarrow +\infty),
\end{equation}
which is the large $n$ behavior of $\E [M(n)]$ found by Comtet and Majumdar in the equation (49) of Ref. \cite{AS2005}, as expected. It is important to notice that the leading term on the right-hand side of Eq.\ (\ref{eq4.8}) gives the leading large $n$ behavior of $\E[M(n)]$ correctly for all $\nu>\mu$ [see Eq. (\ref{eq1.3})]. It is the next subleading correction which needs $\nu>\mu +1$ to be a constant. In the complementary domain $\mu<\nu<\mu +1$, one finds that $\gamma$ in Eq.\ (\ref{eq4.8}) must be replaced with a term growing as $n^{1-(\nu -1)/\mu}$\ \cite{GLM17}.

We conclude this section by noting that, for $c=0$, the exact formula\ (\ref{exact_E}) reduces to
\begin{eqnarray}\label{exact_c0}
\E[M(n)] = \sum_{m=1}^n m^{1/\mu-1} \int_0^{+\infty}
x \,{\mathcal F}_{\mu ,m}\left(x\right)\, dx \;,
\end{eqnarray}
where ${\mathcal F}_{\mu ,m}$ is the PDF of $m^{-1/\mu} \sum_{i=1}^m \eta_i = y_m\, m^{-1/\mu}$ where $y_m$ is the position of the random walker 
in the absence of drift, {i.e.} evolving via $y_i = y_{i-1} + \eta_i$ starting from $y_0 = 0$, after $m$ steps. Therefore, one has $\int_0^{+\infty} dx \, x \,{\mathcal F}_{\mu ,m}\left(x\right) = m^{-1/\mu} \langle |y_m| \rangle/2$ and finally Eq. (\ref{exact_c0}) can be simply written as (see also \cite{dembo})
\begin{eqnarray}\label{exact_c0_final}
\E[M(n)] = \frac{1}{2}\sum_{m=1}^n \frac{\langle |y_m|\rangle}{m}  \;,
\end{eqnarray} 
which is an exact formula, valid for any $n$ and any symmetric and continuous jump distribution. 
%
%
\section{Large $\bm{n}$ behavior of $\bm{\E[M_c(n)]}$ for biased discrete time random walks ($\bm{c\ne 0}$)}\label{sec5}
Henceforth, we will restrict ourselves to the wide class of jump distributions defined by Eq.\ (\ref{eq1.3}) with $\nu\ge 2$. We now turn to the large $n$ behavior of $\E[M_c(n)]$ for a random walk with a drift ($c\ne 0$). To this end, we need the behavior of the generating function in Eq.\ (\ref{eq3.6b}) near $s=1$, hence the one of $\Psi^{(c)}(s)$. We shall mainly use the power series representation\ (\ref{eq3.8}) of $\Psi^{(c)}(s)$. Since all the ${\cal F}_{\mu ,m}$s are symmetric and normalized to unity, one can rewrite\ (\ref{eq3.8}) as
\begin{equation}\label{eq5.1}
\Psi^{(c)}(s)=\frac{c\,\theta(c)s}{1-s}+\Psi^{(-\vert c\vert)}(s),
\end{equation}
with 
\begin{eqnarray}\label{eq5.2}
\Psi^{(-\vert c\vert)}(s)&=&\sum_{m=1}^{+\infty}\frac{s^m}{m^{1-1/\mu}}\int_0^{+\infty}
x{\mathcal F}_{\mu ,m}\left(x+m^{1-1/\mu}\vert c\vert\right)\, dx \nonumber \\
&=&\sum_{m=1}^{+\infty} s^m\int_{\vert c\vert}^{+\infty}
\left(x-\vert c\vert\right)m^{1-1/\mu}{\mathcal F}_{\mu ,m}\left(m^{1-1/\mu}x\right)\, dx,
\end{eqnarray}
where we have made the change of variable $x\to m^{1-1/\mu}(x-\vert c\vert)$. Thus, the analytical properties of $\Psi^{(-\vert c\vert)}(s)$ and its behavior near its dominant singularity depend on the large $m$ behavior of $m^{1-1/\mu}{\mathcal F}_{\mu ,m}(m^{1-1/\mu}x)$. From the definition\ (\ref{eq2.2.3a}) of ${\mathcal F}_{\mu ,m}(x)$ in which one makes the change of variable $q=k/m^{1-1/\mu}$, one gets
\begin{equation}\label{eq5.3}
m^{1-1/\mu}{\cal F}_{\mu ,m}\left(m^{1-1/\mu}x\right)=
\int_{-\infty}^{+\infty}\hat{f}\left(\frac{k}{m}\right)^m
{\rm e}^{-ikx}\, \frac{dk}{2\pi},
\end{equation}
which is nothing but the PDF of the sample mean $m^{-1}\sum_{i=1}^m \eta_i$. It follows that for $x>\vert c\vert$, the large $m$ behavior of $m^{1-1/\mu}{\mathcal F}_{\mu ,m}(m^{1-1/\mu}x)$ is given by a large deviation principle\ \cite{Ell} which makes it possible to determine the leading term of the large $n$ behavior of $\E[M_c(n)]$, as we will now see.
%
%
\subsection{$\bm{1<\mu <2}$: leading behavior of $\bm{\E[M_c(n)]-c\,\theta(c) n}$ in the large $\bm{n}$ limit}\label{sec5.1}
First, we consider a L\'evy flight of index $1<\mu<2$. In this case, the appropriate large deviation principle is the Nagaev's theorem\ \cite{Nag} according to which, away from $x=0$,
\begin{equation}\label{eq5.1.1.1}
n^{1-1/\mu}{\cal F}_{\mu ,n}\left(n^{1-1/\mu}x\right)\sim n^2 f_{tail}(nx)
=\frac{\Gamma(\mu +1)}{\pi an^{\mu -1}}\sin\left(\frac{\pi\mu}{2}\right)
\left(\frac{a}{\vert x\vert}\right)^{\mu +1}\ \ \ \ \ (n\to +\infty),
\end{equation}
where $f_{tail}(\eta)$ denotes the tail of the jump distribution $f(\eta)$ for large values of $\eta$. (The most general formulation of this result can be found in\ \cite{DDS2008}). Equation\ (\ref{eq5.1.1.1}) means that for a heavy-tailed jump distribution, a large deviation $x$ is much more likely to be achieved by a single giant  leap of size $nx$ rather than by $\sim n$ steps of size $\sim x$. Injecting the large $n$ behavior\ (\ref{eq5.1.1.1}) into\ (\ref{eq5.2}) yields the near $s=1$ behavior
\begin{equation}\label{eq5.1.1.2}
\Psi^{(-\vert c\vert)}(s)\sim\vert c\vert\, C\frac{\Gamma(2-\mu)}{(1-s)^{2-\mu}}
\ \ \ \ \ (s\to 1),
\end{equation}
with
\begin{equation}\label{eq5.1.1.3}
C=\frac{\Gamma(\mu -1)}{\pi}\sin\left(\frac{\pi\mu}{2}\right)\left(\frac{a}{\vert c\vert}\right)^\mu .
\end{equation}
The near $s=1$ behavior of the generating function\ (\ref{eq3.6b}) is then obtained from\ (\ref{eq5.1}) and\ (\ref{eq5.1.1.2}). One finds
\begin{equation}\label{eq5.1.1.4}
\sum_{n\ge 0}\E[M_c(n)]s^n\sim\frac{c\,\theta(c) s}{(1-s)^2}
+\vert c\vert\, C\frac{\Gamma(2-\mu)}{(1-s)^{3-\mu}}
\ \ \ \ \ (s\to 1),
\end{equation}
and by Darboux's theorem\ \cite{Hen},
\begin{equation}\label{eq5.1.1.5}
\frac{\E[M_c(n)]-c\,\theta(c) n}{\vert c\vert}\sim\frac{Cn^{2-\mu}}{2-\mu}\ \ \ \ \ (n\to +\infty).
\end{equation}
Equation\ (\ref{eq5.1.1.5}) gives the leading term of the large $n$ behavior of $\E[M_c(n)]$ only. To get the subdominant terms that may survive the limit $n\to +\infty$, we need to determine the corrections to the Nagaev's theorem\ (\ref{eq5.1.1.1}), which is a highly non trivial task. For the whole class of jump distributions defined in Eq.\ (\ref{eq1.3}) with $\nu\ge 2$, the analytical tools developed in Sec.\ \ref{sec3} make it possible to obtain all these surviving terms. This is the subject of the next subsection.
%
%
\subsection{$\bm{1<\mu <2}$: surviving subleading terms of $\bm{\E[M_c(n)]}$ in the large $\bm{n}$ limit}\label{sec5.2}
The calculation goes in two steps. First, we consider the case of a stable jump distribution of index $\mu$, where the surviving terms of the large $n$ expansion of $\E[M_c(n)]$ can be computed explicitly. Then, we show that the result can be extended to all the jump distributions in the class defined by Eq.\ (\ref{eq1.3}), with the same $\mu$, to within a mere shift of the constant term.
\subsubsection{Stable jump distribution with $1<\mu <2$}\label{sec5.2.1}
To begin with, assume that the jumps are drawn from the stable law of index $\mu$. In this case, one has ${\cal F}_{\mu ,n}(x)= a^{-1}f_{S,\mu}(x/a)$ for all $n\ge 1$ and Eq.\ (\ref{eq5.2}) reads
\begin{equation}\label{eq5.1.2.1}
\Psi_S^{(-\vert c\vert)}(s)=\sum_{l=1}^{+\infty} s^l\int_{\vert c\vert}^{+\infty}
\left(x-\vert c\vert\right)
\frac{l^{1-1/\mu}}{a}f_{S,\mu}\left(l^{1-1/\mu}\frac{x}{a}\right)\, dx,
\end{equation}
where the subscript $S$ refers to the stable jump distribution and $f_{S,\mu}(x)$ is the normalized stable law with Fourier transform $\hat{f}_{S,\mu}(k)=\exp(-\vert k\vert^\mu)$. Since the integrals in\ (\ref{eq5.1.2.1}) are over $x\ge \vert c\vert >0$ we can replace $f_{S,\mu}(x)$ with its power series representation away from $x=0$,
\begin{equation}\label{eq5.1.2.2}
f_{S,\mu}(x)=\frac{1}{\pi}\sum_{m=1}^{+\infty}(-1)^{m+1}\sin\left(\frac{m\pi\mu}{2}\right)
\frac{\Gamma(m\mu +1)}{m!}\frac{1}{x^{m\mu +1}}
\ \ \ \ \ (x>0),
\end{equation}
the first term of which corresponds to the large deviation principle\ (\ref{eq5.1.1.1}). The higher order terms give the correction to\ (\ref{eq5.1.1.1}) explicitly as a series. The singular part of $\Psi_S^{(-\vert c\vert)}(s)$ at $s=1$ comes from the contribution of the first $\lbrack 1/(\mu -1)\rbrack$ terms in\ (\ref{eq5.1.2.2}), where $\lbrack 1/(\mu -1)\rbrack$ denotes the integer part of $1/(\mu -1)$. The contribution to\ (\ref{eq5.1.2.1}) of the rest of the series in\ (\ref{eq5.1.2.2}), with $m>\lbrack 1/(\mu -1)\rbrack$, yields a regular function of $s$ at $s=1$ denoted in the following by $\Psi_{R,S}^{(-\vert c\vert)}(s)$. After some straightforward algebra, one gets
\begin{equation}\label{eq5.1.2.3}
\Psi_S^{(-\vert c\vert)}(s)=
\frac{\vert c\vert}{\pi}\sum_{m=1}^{\lbrack 1/(\mu -1)\rbrack}(-1)^{m+1}\sin\left(\frac{m\pi\mu}{2}\right)
\frac{\Gamma(m\mu -1)}{m!}\left(\frac{a}{\vert c\vert}\right)^{m\mu}{\rm Li}_{m(\mu -1)}(s)
+\Psi_{R,S}^{(-\vert c\vert)}(s),
\end{equation}
whose behavior near $s=1$ follows from the one of the polylogarithm functions ${\rm Li}_{m(\mu -1)}(s)$ on the right-hand side. There are two possibilities: either $\mu\ne 1+1/p$ for any integer $p$, or $\mu= 1+1/p$ for some integer $p$.
%
%
\bigskip
\paragraph{$\mu\ne 1+1/p$ for any integer $p$.}\label{sec5.1.2.1}
Using Eq.\ (\ref{eq4.2}) in which $1-1/\mu$ is replaced with $m(\mu -1)$, one finds
\begin{equation}\label{eq5.1.2.4}
\Psi_S^{(-\vert c\vert)}(s)=
\vert c\vert\sum_{m=1}^{\lbrack 1/(\mu -1)\rbrack} C_m
\frac{\Gamma(1-m(\mu -1))}{(1-s)^{1-m(\mu -1)}}
+\vert c\vert\kappa_c
+O((1-s)^{\mu -1})\ \ \ \ \ (s\rightarrow 1),
\end{equation}
with
\begin{equation}\label{eq5.1.2.5}
C_m=(-1)^{m+1}\sin\left(\frac{m\pi\mu}{2}\right)
\frac{\Gamma(m\mu -1)}{\pi m!}\left(\frac{a}{\vert c\vert}\right)^{m\mu},
\end{equation}
and
\begin{equation}\label{eq5.1.2.6}
\kappa_c =\sum_{m=1}^{\lbrack 1/(\mu -1)\rbrack} C_m \zeta(m(\mu -1))
+\frac{\Psi_{R,S}^{(-\vert c\vert)}(1)}{\vert c\vert}.
\end{equation}
Note that the constant $C_1$ is precisely the same as $C$ in Eq. (\ref{const1}). 

From\ (\ref{eq3.6b}),\ (\ref{eq5.1}), and\ (\ref{eq5.1.2.4}) one gets
\begin{eqnarray}\label{eq5.1.2.7}
\sum_{n\ge 0}\E[M_{c,S}(n)]s^n&=&\frac{c\sigma(c)s}{(1-s)^2}
+\vert c\vert\sum_{m=1}^{\lbrack 1/(\mu -1)\rbrack} C_m
\frac{\Gamma(1-m(\mu -1))}{(1-s)^{2-m(\mu -1)}} \nonumber \\
&+&\frac{\vert c\vert\kappa_c}{1-s}
+O\left(\frac{1}{(1-s)^{2-\mu}}\right)
\ \ \ \ \ (s\rightarrow 1),
\end{eqnarray}
which translates into the large $n$ behavior\ \cite{Hen}
\begin{equation}\label{eq5.1.2.8}
\frac{\E[M_{c,S}(n)]-c\sigma(c)n}{\vert c\vert}=
\sum_{m=1}^{\lbrack 1/(\mu -1)\rbrack}\frac{C_mn^{1-m(\mu -1)}}{1-m(\mu -1)}
+\kappa_c +O\left(\frac{1}{n^{\mu -1}}\right)\ \ \ \ \ (n\rightarrow +\infty).
\end{equation}
%
%
\paragraph{$\mu= 1+1/p$ for some integer $p$.}\label{sec5.1.2.2}
On the right-hand side of\ (\ref{eq5.1.2.3}) and for $m\ne p$, we use the equation\ (\ref{eq4.2}) in which $1-1/\mu$ is replaced with $m(\mu -1)=m/p$. For $m=p$ we simply use the identity ${\rm Li}_1(s)=\ln\lbrack 1/(1-s)\rbrack$. One obtains
\begin{eqnarray}\label{eq5.1.2.9}
\Psi_S^{(-\vert c\vert)}(s)&=&
\vert c\vert\sum_{m=1}^{p-1} C_m \frac{\Gamma(1-m/p)}{(1-s)^{1-m/p}}
+\vert c\vert C_p\ln\left(\frac{1}{1-s}\right) \nonumber \\
&+&\vert c\vert\kappa_c^\prime
+O((1-s)^{1/p})\ \ \ \ \ (s\rightarrow 1),
\end{eqnarray}
with $C_m$ given by Eq.\ (\ref{eq5.1.2.5}) and
\begin{equation}\label{eq5.1.2.10}
\kappa_c^\prime =\sum_{m=1}^{p-1} C_m \zeta(m/p)
+\frac{\Psi_{R,S}^{(-\vert c\vert)}(1)}{\vert c\vert}.
\end{equation}
Note that, according to\ (\ref{eq5.1.2.5}), $C_p=0$ if $p$ is odd and there is no logarithmic correction in this case. The equations\ (\ref{eq3.6b}),\ (\ref{eq5.1}), and\ (\ref{eq5.1.2.9}) yield
\begin{eqnarray}\label{eq5.1.2.11}
&&\sum_{n\ge 0}\E[M_{c,S}(n)]s^n=\frac{c\sigma(c)s}{(1-s)^2}
+\vert c\vert\sum_{m=1}^{p-1} C_m
\frac{\Gamma(1-m/p)}{(1-s)^{2-m/p}} \nonumber \\
&&+\frac{\vert c\vert C_p}{1-s}\ln\left(\frac{1}{1-s}\right)
+\frac{\vert c\vert\kappa_c^\prime}{1-s}
+O\left(\frac{1}{(1-s)^{1-1/p}}\right)\ \ \ \ \ (s\rightarrow 1).
\end{eqnarray}
The contributions of the algebraic singularities on the right-hand side of\ (\ref{eq5.1.2.11}) to the large $n$ behavior of $\E[M_{c,S}(n)]$ are given by the classical Darboux's theorem\ \cite{Hen}. To get the contribution of the remaining logarithmic singularity we use the result that a factor $(1-s)^{-1}\ln\lbrack 1/(1-s)\rbrack$, ($s\to 1$), in a generating function translates into a factor $\ln n +\gamma_E$, ($n\to +\infty$), in its coefficients, where $\gamma_E$ is the Euler constant\ \cite{FS} (chap. VI, Figure VI.5). After some straightforward algebra, one obtains
\begin{equation}\label{eq5.1.2.12}
\frac{\E[M_{c,S}(n)]-c\sigma(c)n}{\vert c\vert}=
\sum_{m=1}^{p-1}\frac{p\, C_m n^{1-m/p}}{p-m}+C_p\ln n
+\kappa_c +O\left(\frac{1}{n^{1/p}}\right)\ \ \ \ \ (n\rightarrow +\infty),
\end{equation}
with, according to Eq.\ (\ref{eq5.1.2.10}),
\begin{equation}\label{eq5.1.2.13}
\kappa_c =\kappa_c^\prime +C_p\gamma_E
=\sum_{m=1}^{p-1} C_m \zeta(m/p)+C_p\gamma_E
+\frac{\Psi_{R,S}^{(-\vert c\vert)}(1)}{\vert c\vert},
\end{equation}
and $C_p=0$ if $p$ is odd. Note that the leading term in the equations\ (\ref{eq5.1.2.8}) and\ (\ref{eq5.1.2.12}) coincides with the asymptotics\ (\ref{eq5.1.1.5}), as it should be.
%
%
\subsubsection{Generalization to the whole class of jump PDFs defined by Equation\ (\ref{eq1.3}) with $1<\mu <2$}\label{sec5.2.2}
So far, the asymptotic expansions\ (\ref{eq5.1.2.8}) and\ (\ref{eq5.1.2.12}) hold for a stable jump distribution only. We now show that they are actually valid for the whole class of jump distributions defined in Eq.\ (\ref{eq1.3}), with $1<\mu <2$ and $\nu\ge 2$, to within a shift of the constant term $\kappa_c$.

Consider a RW\ (\ref{eq1.1}) with a jump PDF whose Fourier transform behaves as in Eq.\ (\ref{eq1.3}) with $1<\mu <2$ and $\nu\ge 2$. We want to compare the associated expected maximum $\E[M_c(n)]$ with its counterpart $\E[M_{c,S}(n)]$ for the stable jump distribution of same index $\mu$ and same characteristic length $a$, in the large $n$ limit. From the expression\ (\ref{eq3.6b}) of the generating function of $\E[M_c(n)]$ and the integral representation\ (\ref{eq3.7}) of $\Psi^{(c)}(s)$ [together with Eq.\ (\ref{eq5.1})], one has
\begin{equation}\label{general1}
\sum_{n\ge 0}\left(\E[M_c(n)]-\E[M_{c,S}(n)]\right) s^n =\frac{\Psi^{(-|c|)}(s)-\Psi_S^{(-|c|)}(s)}{1-s}
\sim\frac{\vert c\vert\Delta\kappa_c}{1-s}\ \ \ \ \ (s\to 1),
\end{equation}
where
\begin{eqnarray}\label{general2}
\Delta\kappa_c&=&\frac{1}{2\pi\vert c\vert}
 \int_{-\infty}^{+\infty} \ln \left(\frac{1 - \e^{-i |c|q} \hat f(q)}{1 - \e^{-i |c|q - |a\,q|^\mu}} \right)\, \frac{dq}{q^2} \nonumber \\
&=&\frac{1}{2\pi\vert c\vert} \int_{0}^{+\infty} \ln\left(\frac{1 - 2 \cos(cq) \hat f(q) + \hat f(q)^2}{1 - 2 \cos(cq) \e^{-|a\,q|^\mu} + \e^{-2 |a\,q|^\mu}} \right)\, \frac{dq}{q^2} \;.
\end{eqnarray}
Note that $\Delta\kappa_c$ is independent of the sign of $c$. In the small $q$ limit, the integrand in the second line of Eq.\ (\ref{general2}) behaves like $\propto |q|^{\nu + \mu -4}$, which is integrable at $q=0$ if $\mu + \nu > 3$. Thus, the latter condition is clearly satisfied for 
$1< \mu  \leq 2$ and $\nu\ge 2$, that we are considering here. Moreover, since $\hat f(q) \to 0$ in the large $q$ limit, the integrand in the second line of Eq.\ (\ref{general2}) is also integrable at $q\to +\infty$. Thus, $\Delta\kappa_c <+\infty$ for the class of jump distributions defined in Eq.\ (\ref{eq1.3}) with $1 < \mu <2$ and $\nu\ge 2$, and it follows immediately from Darboux's theorem\ \cite{Hen} and the simple pole behavior near $s=1$ in Eq.\ (\ref{general1}) that
\begin{equation}\label{general3}
\lim_{n\to +\infty}\E[M_c(n)]-\E[M_{c,S}(n)] =\vert c\vert\Delta\kappa_c .
\end{equation}
Finally, it remains to replace $\E[M_{c,S}(n)]$ in Eq.\ (\ref{general3}) with the appropriate asymptotic expansion\ (\ref{eq5.1.2.8}) or\ (\ref{eq5.1.2.12}), and one finds that the large $n$ asymptotic expansion of $\E[M_c(n)]$ is {\it exactly} the same as for the stable jump distribution of same $\mu$ and $a$, up to the constant term which is shifted from $\kappa_c$ to $\kappa_c +\Delta\kappa_c$, where $\kappa_c$ and $\Delta\kappa_c$ are respectively given in Eqs.~(\ref{eq5.1.2.6}), (\ref{eq5.1.2.13}), and\ (\ref{general2}).

We now turn to the last remaining case of biased, Brownian-like, random walks with index $\mu =2$.
%
%
\subsection{$\bm{\mu =2}$: large $\bm{n}$ behavior of $\bm{\E[M_c(n)]}$}\label{sec5.3}
In this case, the jumps have a finite variance and the appropriate large deviation principle to be used in Eq.\ (\ref{eq5.2}) is the Cram\'er's theorem\ \cite{Cram} according to which there is a positive and convex continuous function $I(x)$, called the rate function, with a single minimum at $x=0$ (with $I(0)=0$) such that
\begin{equation}\label{eq5.2.1}
\sqrt{n}\, {\cal F}_{2,n}\left(\sqrt{n}\, x\right)\sim
h(n){\rm e}^{-nI(x)}\ \ \ \ \ (n\rightarrow +\infty),
\end{equation}
where the prefactor $h(n)$ is subdominant with respect to the decaying exponential. Both $I(x)$ and $h(n)$ are sensitive to the jump distribution $f(\eta)$ and there is no generic expressions for these two functions. Injecting\ (\ref{eq5.2.1}) into the generating function on the right-hand side of\ (\ref{eq5.2}) with $\mu =2$ and performing the integration over $x$ for large $n$, one finds that the coefficients behave asymptotically as
\begin{equation}\label{eq5.2.2}
\sqrt{n}\, \int_{\vert c\vert}^{+\infty}
\left(x-\vert c\vert\right){\mathcal F}_{2,n}\left(\sqrt{n}\, x\right)\, dx\sim
\frac{1}{\left(I^\prime \vert c\vert\right)^2}\, \frac{h(n)}{n^2}
{\rm e}^{-nI\left(\vert c\vert\right)}\ \ \ \ \ (n\rightarrow +\infty).
\end{equation}
Because of the exponential on the right-hand side of\ (\ref{eq5.2.2}), the dominant singularity of $\Psi^{(-\vert c\vert)}(s)$ is located at $s=\exp I(\vert c\vert) >1$ from which it follows that $\Psi^{(-\vert c\vert)}(s)$ is analytic at $s=1$. Writing
\begin{equation}\label{eq5.2.3}
\sum_{n\ge 0}\E[M_c(n)]\,s^n =\frac{c\,\theta(c)s}{(1-s)^2}+\frac{\Psi^{(-\vert c\vert)}(s)}{1-s},
\end{equation}
where we have used\ (\ref{eq5.1}), it is clear that the large $n$ behavior of $\E[M_c(n)]$ is given by the pole at $s=1$ with residue $\Psi^{(-\vert c\vert)}(1)$. The singularity of $\Psi^{(-\vert c\vert)}(s)$ at $s=\exp I(\vert c\vert) >1$ determines the subdominant, vanishing, correction to this behavior. Applying the Darboux's theorem\ \cite{Hen} for each of these two singularities, one obtains
\begin{equation}\label{eq5.2.4}
\frac{\E[M_c(n)]-c\,\theta(c)n}{\vert c\vert}=\kappa_c
+O\left(\frac{h(n)}{n^2}
{\rm e}^{-nI\left(\vert c\vert\right)}\right)
\ \ \ \ \ (n\rightarrow +\infty),
\end{equation}
with $\kappa_c =\Psi^{(-\vert c\vert)}(1)/\vert c\vert$. According to the expressions\ (\ref{eq3.7}) and\ \ (\ref{eq3.8}) for $\Psi^{(c)}(s)$, the integral and power series representations of $\kappa_c$ are respectively given by
\begin{equation}\label{eq5.2.5}
\kappa_c =\frac{1}{2\pi}\frac{\partial}{\partial\lambda}\left.\int_{-\infty}^{+\infty}
\frac{\ln\lbrack 1-\hat{f}(q/c){\rm e}^{-iq}\rbrack}{\lambda +iq}\, dq\right\vert_{\lambda =0},
\end{equation}
where $\lambda$ goes to zero from ${\rm Re}(\lambda)=+\infty$, and
\begin{equation}\label{eq5.2.6}
\kappa_c=\frac{a}{\vert c\vert}\sum_{m=1}^{+\infty}\frac{1}{\sqrt{m}}
\int_0^{+\infty}x{\cal F}_{2,m}^{\rm\, resc}\left(x+\sqrt{m}\, \frac{\vert c\vert}{a}\right)\, dx,
\end{equation}
where we have made the change of variable $x\to ax$ and introduced the rescaled function ${\cal F}_{2,m}^{\rm\, resc}(x)=a{\cal F}_{2,m}(ax)$ in order to make the dependence on $a$ and $c$ more explicit. In principle, Eqs.\ (\ref{eq5.2.5}) and\ (\ref{eq5.2.6}) make it possible to compute $\kappa_c$ for any jump distribution with a finite variance. For instance, for a Gaussian jump distribution one has ${\cal F}_{2,m}^{\rm\, resc}(x)=1/(2\sqrt{\pi})\,\e^{-x^2/4}$ for any $m\ge 1$ and Eq.\ (\ref{eq5.2.6}) gives $\kappa_c$ as the series
\begin{eqnarray}\label{eq5.2.7}
\kappa_c = \sum_{m=1}^\infty \left[\frac{\e^{-b^2\,m}}{2b \sqrt{\pi\,m}}  - \frac{1}{2} {\rm erfc}\left(b\sqrt{m} \right) \right] \;,
\end{eqnarray}
where $b=|c|/2a$, as announced in Eq.\ (\ref{gamma_gaussian}).

Although $\kappa_c$ in Eqs.\ (\ref{eq5.2.5}) or\ (\ref{eq5.2.6}) does depend on the jump distribution explicitly, its small $c$ behavior does not, as we will now see. This is most easily shown by using the power series representation\ (\ref{eq5.2.6}). In the limit $c\to 0$, the sum over $m$ on the right-hand side of\ (\ref{eq5.2.6}) is a Riemann sum which can be replaced with a integral over $y=\sqrt{m}\, |c|/a$. One gets
\begin{eqnarray}\label{eq5.2.8}
\kappa_c&\sim&\frac{2a^2}{|c|^2}\int_0^{+\infty}dy
\int_0^{+\infty}dx\, x{\cal F}_{2,(ya/|c|)^2}^{\rm\, resc}(x+y) \nonumber \\
&\sim&\frac{2a^2}{|c|^2}\int_0^{+\infty}dy
\int_0^{+\infty}dx\, x{\cal F}_{2,\infty}^{\rm\, resc}(x+y)\ \ \ \ \ (c\to 0),
\end{eqnarray}
in which ${\cal F}_{2,\infty}^{\rm\, resc}(x+y)=f_{S,2}(x+y)=1/(2\sqrt{\pi})\,\e^{-(x+y)^2/4}$ [see Eq. (\ref{Fmu_asympt}) with $\mu=2$], and after some straightforward algebra, one finds
\begin{equation}\label{eq5.2.9}
\kappa_c\sim\frac{a^2}{\sqrt{\pi}\, |c|^2}\int_0^{+\infty}dy
\int_0^{+\infty}dx\, x \e^{-(x+y)^2/4}=\frac{a^2}{|c|^2}\ \ \ \ \ (c\to 0),
\end{equation}
independent of the jump distribution (with $\mu =2$).
%
%
\subsection{Crossover between the large $\bm{n}$ behaviors of $\bm{\E[M_c(n)]}$ for $\bm{c=0}$ and a small nonzero $\bm{c}$}\label{sec5.4}
From the results in Sec.\ \ref{sec4} (for $c=0$) and Sec.\ \ref{sec5} (for $c\ne 0$) it is clear that the two limits $n\to\infty$ and $c\to 0$ do not commute, hence the small $c$ limit of the large $n$ behavior of $\E[M_c(n)]$ is singular. This suggests the existence of a scaling regime describing the crossover between the leading large $n$ behaviors of $\E[M_c(n)]$ for $c=0$ and a small nonzero $c$.

To determine the corresponding scaling form, we analyze the behavior of $\E[M_c(n)]$ in the limits $c \to 0$ and $n \to \infty$, keeping the scaling variable $u = |c| n^{1-1/\mu}/a$ fixed. The starting point of our analysis is the exact formula given in Eq. (\ref{exact_E_2}), which we write as
\begin{eqnarray}\label{cross_1}
\frac{\E[M_c(n)] - c \, \theta(c)\, n}{a} = \sum_{m=1}^n m^{1/\mu-1} \int_0^{+\infty} \frac{x}{a} \,{\mathcal F}_{\mu ,m}\left(x + m^{1-1/\mu}|c|\right)\, dx \;.
\end{eqnarray}
In the large $n$ limit, the sum over $m$ is dominated by the large values of $m$. Besides, when $c \to 0$, the sum over $m$ on the right-hand side of Eq.\ (\ref{cross_1}) is a Riemann sum which can be replaced with an integral over $y = m^{1-1/\mu}|c|$. This yields
\begin{eqnarray}\label{cross_2}
&&\frac{\E[M_c(n)] - c \, \theta(c)\, n}{a} \sim \nonumber \\
&&\frac{\mu}{\mu-1}|c|^{-\frac{1}{\mu-1}} \int_0^{|c|\,n^{1-1/\mu}} dy\, y^{\frac{2-\mu}{\mu-1}} \int_0^{+\infty} dx \, \frac{x}{a} \, {\cal F}_{\mu, \left(y/|c|\right)^{\frac{\mu}{\mu-1}}}(x+y)  \;.
\end{eqnarray}
In the limit $c \to 0$, we can replace ${\cal F}_{\mu, \left(y/|c|\right)^{\frac{\mu}{\mu-1}}}(x+y)$ by its limiting behavior ${\cal F}_{\mu, \infty}(x+y) = a^{-1}f_{S,\mu}((x+y)/a)$ [see Eq. (\ref{Fmu_asympt})] and by performing the changes of variables $x \to x/a$ and $y \to y/a$ one finally arrives at
\begin{equation}\label{cross_3}
\frac{\E[M_c(n)] - c \, \theta(c)n}{a} \sim n^{1/\mu} {\cal G}_\mu\left(\frac{|c|}{a} n^{1-1/\mu} \right) \:,
\end{equation}
with
\begin{equation}\label{cross_4}
{\cal G}_\mu(u) = \frac{\mu}{\mu-1} u^{-\frac{1}{\mu-1}} \int_0^u dy \, y^{\frac{2-\mu}{\mu-1}} \int_0^{+\infty} dx \, x\, \, f_{S,\mu}(x+y) \;,
\end{equation}
which is indeed the scaling form given in Eqs.\ (\ref{scaling_mu_1}) and\ (\ref{scaling_mu}) in the introduction. For $\mu =2$, the stable law is the Gaussian distribution  and one has $f_{S,2}(x+y)=1/(2\sqrt{\pi})\,\e^{-(x+y)^2/4}$. The right-hand side of\ (\ref{cross_4}) can then be computed explicitly, yielding
\begin{equation}\label{cross_5}
{\cal G}_2(u) = \frac{\e^{-\frac{u^2}{4}}}{\sqrt{\pi}} + \frac{{\rm erf}{\left(\frac{u}{2}\right)}}{u} - \frac{1}{2} u\, {\rm erfc}\left(\frac{u}{2}\right) \;,
\end{equation}
with ${\cal G}_2(u) \sim 2/\sqrt{\pi}$ for $u \to 0$ and ${\cal G}_2(u) \sim 1/u$ as $u \to \infty$. For $1<\mu <2$, there is no explicit expression of the scaling function but the small and large argument behaviors can always be obtained. In the small $u$ limit, it follows from Eq.\ (\ref{cross_4}) that ${\cal G}_\mu(u)$ behaves as
\begin{eqnarray}\label{cross_6}
{\cal G}_\mu(u)&\sim&\frac{\mu}{\mu-1} u^{-\frac{1}{\mu-1}} \int_0^u y^{\frac{2-\mu}{\mu-1}}dy
\int_0^{+\infty} x\, f_{S,\mu}(x)\, dx \nonumber \\
&=&\frac{\mu}{\pi}\, \Gamma\left(1-\frac{1}{\mu}\right)\ \ \ \ \ (u\to 0),
\end{eqnarray}
where we have used the identity $\int_0^{+\infty}x\, f_{S,\mu}(x)\, dx=\Gamma(1-1/\mu)/\pi$. In the opposite large $u$ limit, we make the change of variable $y=uz$ on the right-hand side of Eq.\ (\ref{cross_4}) and by using the asymptotic behavior $f_{S,\mu}(\eta)\sim\pi^{-1}\sin(\pi\mu/2)\Gamma(\mu +1)\eta^{-\eta -1}$ ($\eta\to +\infty$), one gets
\begin{eqnarray}\label{cross_7}
{\cal G}_\mu(u)&\sim&\frac{\mu\Gamma(\mu +1)}{\pi(\mu -1)}\sin\left(\frac{\pi\mu}{2}\right)
\int_0^1 z^{\frac{2-\mu}{\mu-1}}dz\int_0^{+\infty}\frac{x\, dx}{(x+uz)^{\mu +1}} \nonumber \\
&=&\frac{\Gamma(\mu -1)}{\pi(2-\mu)}\sin\left(\frac{\pi\mu}{2}\right)\, \frac{1}{u^{\mu -1}}
\ \ \ \ \ (u\to +\infty).
\end{eqnarray}
From these small and large argument behaviors of ${\cal G}_\mu(u)$ it can be checked that the scaling form\ (\ref{cross_3}) reduces to the leading term of\ (\ref{eq4.8}) for $c=0$, and to Eq.\ (\ref{eq5.1.1.5}) for $1<\mu <2$ and $c\ne 0$, or Eq.\ (\ref{eq5.2.4}) [with $\kappa_c$ given in Eq.\ (\ref{eq5.2.9})] for $\mu =2$ and a small nonzero $c$.

Finally, it is worth noticing that, for $1<\mu<2$, the leading behavior on the right-hand side of Eqs.\ (\ref{eq4.8}) and\ (\ref{eq5.1.1.5}), as well as the scaling form\ (\ref{cross_3}), actually hold for all $\nu>\mu$ in Eq.\ (\ref{eq1.3}). Unlike the subleading corrections which need either $\nu>\mu +1$ in Eq.\ (\ref{eq4.8}) or $\nu\ge 2$ in the results of Sec.\ \ref{sec5.2}.
%
%
\section{Numerical simulations}\label{sec6}
We have performed numerical simulations of random walks with a drift, as defined in Eq. (\ref{eq1.1}) with different jump distributions $f(\eta)$. The statistics of the maximum $M_c(n)$ was obtained by sampling $10^7$ independent random walks of $10^3$ steps. In all cases, we have compared the results of these numerical simulations (i) with a numerical evaluation of our exact finite $n$ expression\ (\ref{exact_E}) and (ii) with our analytical asymptotic results, valid for large $n$. All the results presented here are for $c<0$, without loss of generality, according to the identity\ (\ref{identity_av}), $\E[M_c(n)]-cn=\E[M_{-c}(n)]$, that we have checked numerically. We present the cases $\mu = 2$ and $1<\mu<2$ separately.  
%
%
\subsection{$\bm{\mu = 2}$}

\begin{figure}
\includegraphics[width = \linewidth]{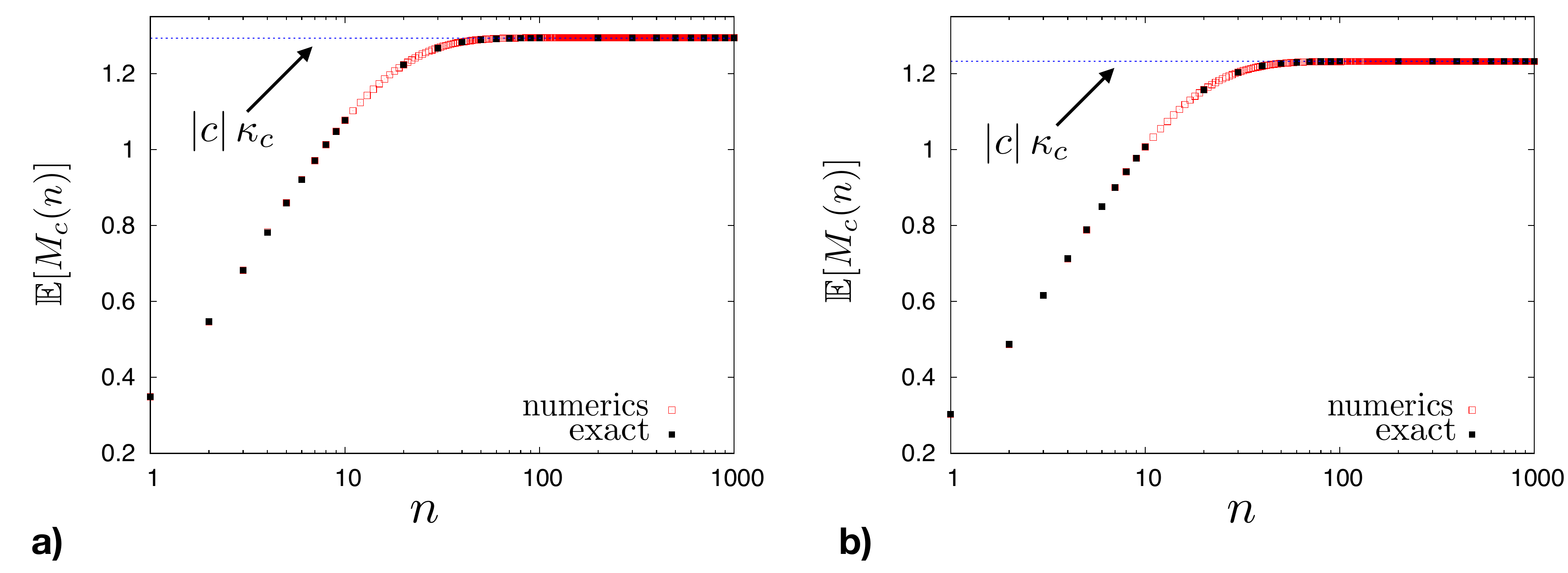}
\caption{{\bf a)} Plot of $\E[M_c(n)]$ for a RW with Gaussian jumps, corresponding to $\mu = 2$, with variance $\sigma = \sqrt{2}$, i.e. $a=1$, and a negative drift $c=-1/2$, as a function of $n$ (in a log-scale). The red squares correspond to numerical simulations of the RW while the black squares are obtained by a numerical evaluation of the exact formula in Eq. (\ref{exact_E}). The blue dotted line corresponds to the exact asymptotic value $|c|\,\kappa_c = 1.29376\ldots$ given in Eq. (\ref{eq5.2.7}). {\bf b)} Plot of $\E[M_c(n)]$ for a RW with (symmetric) exponential
 jumps, corresponding to $\mu = 2$ and $a=1$, and a negative drift $c=-1/2$, as a function of $n$ (in a log-scale). The red squares correspond to numerical simulations of the RW while the black squares are obtained by a numerical evaluation of the exact formula in Eq. (\ref{exact_E}). The blue dotted line corresponds to the exact asymptotic value $|c|\,\kappa_c = 1.23233\ldots$ given by a numerical evaluation of Eq. (\ref{eq5.2.6}).}\label{fig_mu2}
\end{figure}

In Fig. \ref{fig_mu2}a) we show $\E[M_c(n)]$ as obtained from numerical simulations of the RW with a Gaussian jump distribution $f(\eta) = f_{S,2}(\eta)=  \e^{-\eta^2/4}/\sqrt{4 \pi}$, corresponding to $a=1$, and $c=-1/2$ (red squares). We have compared these numerical values with the exact ones given in Eq. (\ref{exact_E}) (black squares), where in this case ${\cal F}_{2,m}(x) = f_{S,2}(x)$, independent of $m$. We can see that the numerical results are in perfect agreement with the expected, exact, values\ (\ref{exact_E}). In Fig. \ref{fig_mu2}a), we have also indicated our analytical prediction of the leading asymptotic behavior of $\E[M_c(n)] $ for large $n$ (blue dotted line).   

In Fig. \ref{fig_mu2}b) we show the same quantities (with the same corresponding symbols) but for symmetric exponential jumps $f(\eta) = (1/2)\,\e^{-|\eta|}$. In this case ${\cal F}_{2,m}(x)$ is a non-trivial function of both $x$ and $m$, and the evaluation of our exact expression (\ref{exact_E}) was carried out by using the equivalent form\ (\ref{exact_E_3}). Again, the agreement between numerical results and analytical predictions is very good. 
%
%
\subsection{$\bm{1 < \mu < 2}$}
\begin{figure}
\includegraphics[width = \linewidth]{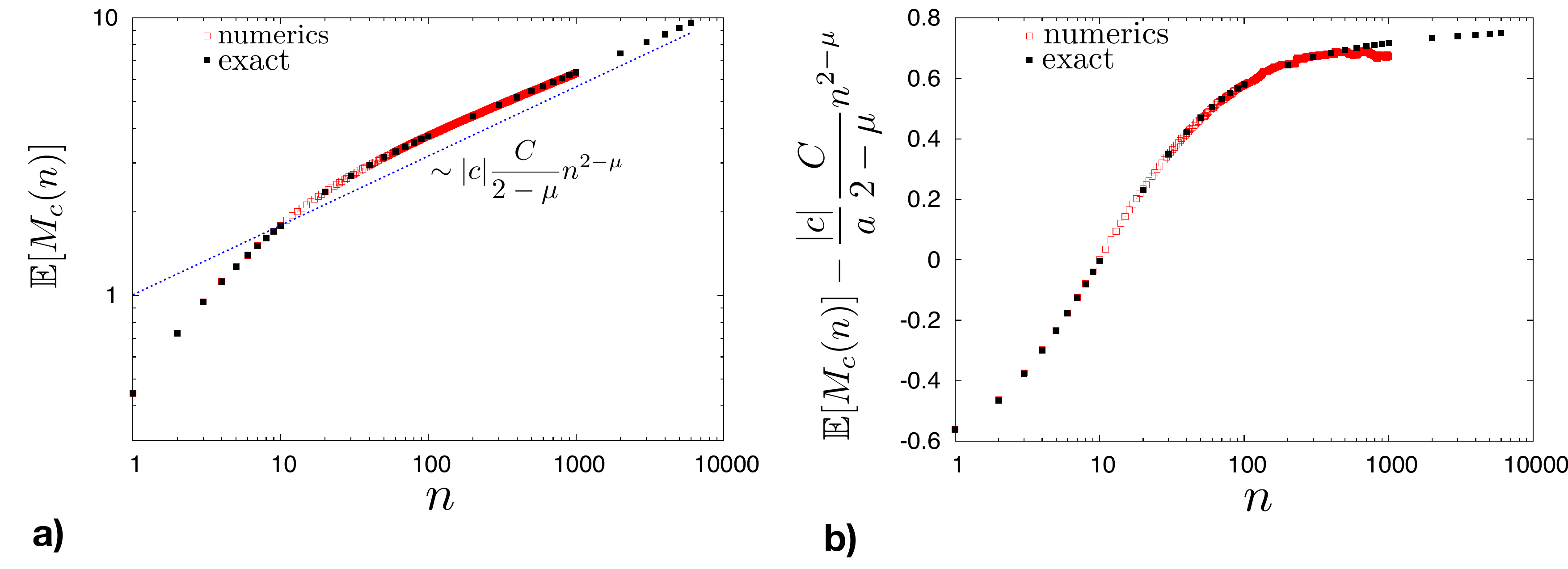}
\caption{{\bf a)} Log-log plot of $\E[M_c(n)]$ for a RW with a stable jump distribution with $\mu = 7/4$, $a=1$ and a negative drift $c=-1/2$ as a function of $n$. The red squares correspond to numerical simulations of the RW while the black squares are obtained by a numerical evaluation of the exact formula in Eq. (\ref{exact_E}). The dotted line corresponds to the leading term for large $n$, $\E[M_c(n)] \sim |c| C/(2-\mu)\,n^{2-\mu}$ [see Eq.~(\ref{eq5.1.1.5})].  {\bf b)} Plot of $\E[M_c(n)] - |c| C/(2-\mu)\,n^{2-\mu}$ as a function of $n$ (in a log-scale), which suggests that the sub-leading term in the expansion of $\E[M_c(n)]$ for large $n$ is a constant, in agreement with the formula in Eq. (\ref{eq5.1.2.8}), since $1- \mu= 3/4 \neq 1/p$ for all integer $p$ in this case.}\label{Fig_mu34}
\end{figure}

In Fig.~\ref{Fig_mu34}, we show the results for a L\'evy flight with $c=-1/2$ and a stable jump distribution of index $\mu = 7/4$ and scale parameter $a=1$. As before, the red and black squares correspond respectively to the numerical estimate [by simulating the RW in Eq. (\ref{eq1.1})] and to the exact value, for finite $n$, obtained from Eq. (\ref{exact_E}) [or the equivalent expression\ (\ref{exact_E_3})]. It can be seen in Fig. \ref{Fig_mu34}a) that the agreement between red and black squares is quite good. The blue dotted line corresponds to the leading asymptotic behavior, $\E[M_c(n)] \sim |c| C/(2-\mu)\,n^{2-\mu}$, given in Eq. (\ref{eq5.1.1.5}). The tendency of the squares to align parallel to the line, but slightly away from it, suggests that the subleading correction is a constant as expected from our result in Eq. (\ref{eq5.1.2.8}) with $\mu =7/4$. This is confirmed by the results in Fig.\ \ref{Fig_mu34}b) suggesting more clearly that the difference $\E[M_c(n)] - |c| C/(2-\mu)\,n^{2-\mu}$ goes to a constant when $n$ gets large.
  
\begin{figure}
\includegraphics[width = \linewidth]{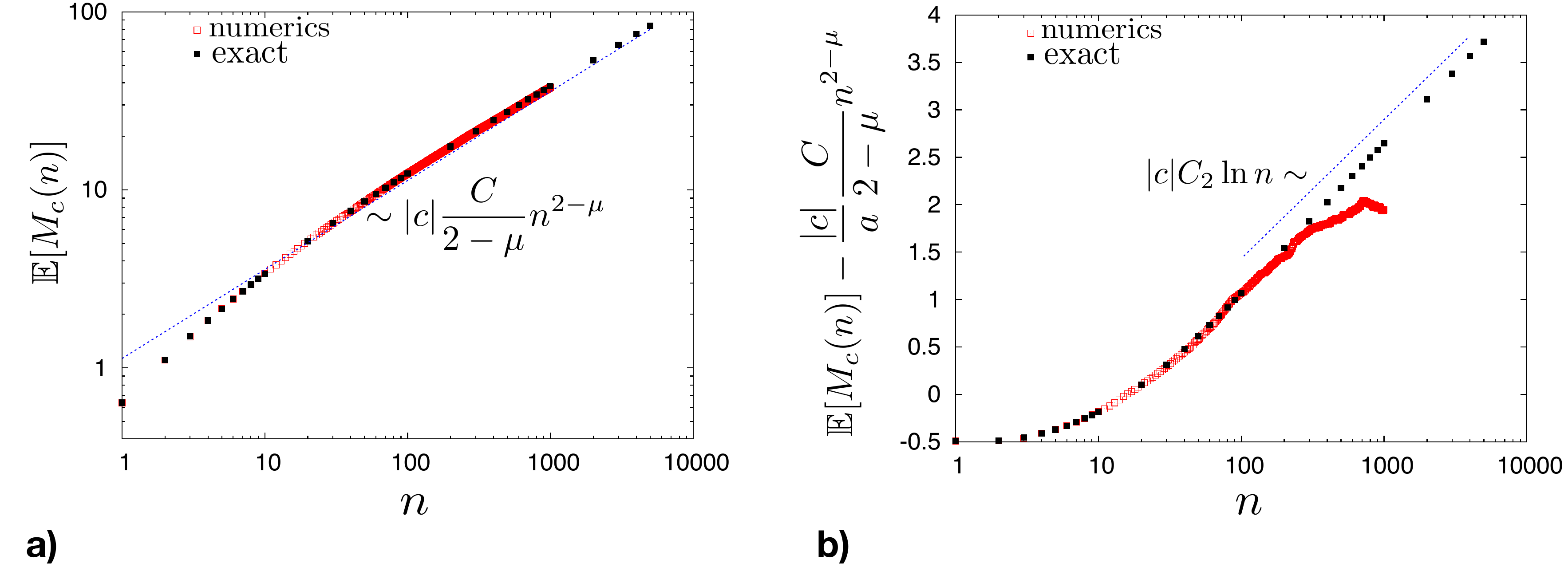}
\caption{{\bf a)} Log-log plot of $\E[M_c(n)]$ for a RW with a stable jump distribution with $\mu = 3/2$, $a=1$ and a negative drift $c=-1/2$, as a function of $n$. The red squares correspond to numerical simulations of the RW while the black squares are obtained by a numerical evaluation of the exact formula in Eq. (\ref{exact_E}). The dotted line corresponds to the leading term for large $n$, $\E[M_c(n)] \sim |c| C/(2-\mu)\,n^{2-\mu}$ [see Eq.~(\ref{eq5.1.1.5})]. {\bf b} Plot of $\E[M_c(n)] - |c| C/(2-\mu)\,n^{2-\mu}$ as a function of $n$ (in a log-scale), which suggests that the sub-leading term in the expansion of $\E[M_c(n)]$ for large $n$ is $\propto |c| C_2 \ln n$ (as indicated by the dotted line), in agreement with the formula in Eq. (\ref{eq5.1.2.12}), since $1- \mu= 1/2$ in this case.}\label{Fig_mu12}
\end{figure}

In Fig. \ref{Fig_mu12} we show the results for a L\'evy flight with a stable jump distribution of index $\mu = 3/2$, all other parameters equal. In this case, our results in Sec.\ \ref{sec5.1.2.2} predict that the first subleading term is a logarithmic correction, $|c|C_2\ln n$, and the next one is a constant $\kappa_c$. Fig. \ref{Fig_mu12}a) shows that the leading behavior $\E[M_c(n)] \sim |c| C/(2-\mu)\,n^{2-\mu}$ already describes the data quite well for $n$ large enough. In Fig. \ref{Fig_mu12}b) we have subtracted this leading term and the blue dotted line corresponds to the first, logarithmic, correction. The tendency of the squares to align parallel to this line, but slightly away from it (the difference corresponding to the constant term $\kappa_c$), is in good agreement with our prediction in Eq. (\ref{eq5.1.2.12}) for $\mu= 3/2$. Note that the logarithmic correction is clearly seen on the exact results (black squares) which allow to reach higher values of $n$, while the numerical data gets too noisy (red squares).       

\begin{figure}[ht]
\includegraphics[width=\linewidth]{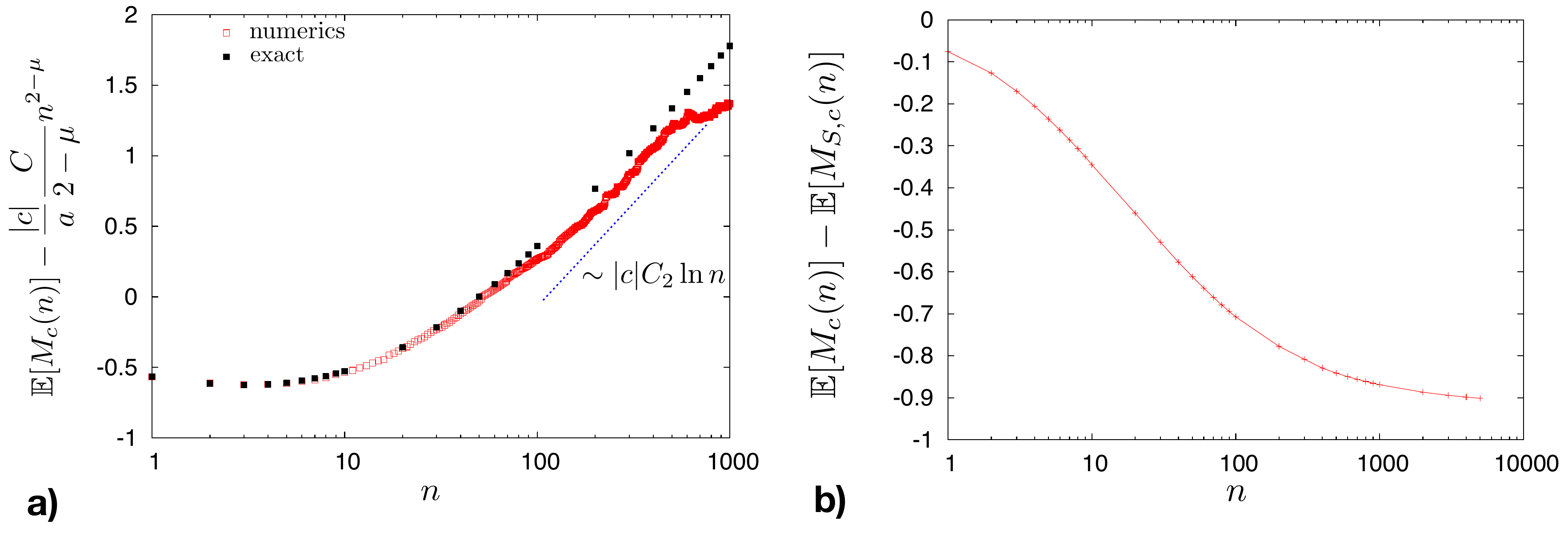}
\caption{{\bf a)} Plot of $\E[M_c(n)] - |c| \frac{C}{2-\mu} n^{2-\mu}$ as a function of $n$ (on a log-scale) for a Pareto jump distribution with parameters $\mu =3/2$,  $a=1$, and a negative drift $c=-1/2$. We recall that $|c| \frac{C}{2-\mu} n^{2-\mu}$ is the leading large $n$ behavior (\ref{eq5.1.1.5}), while the first correction is expected to be $|c|\, C_2 \ln n$ [see Eq. (\ref{eq5.1.2.12})], which is indicated by the blue dotted line. The red squares correspond to numerical simulations of the RW (\ref{eq1.1}) while the black squares represent the exact formula (\ref{exact_E_2_app3}). {\bf b)} Plot of the difference $\E[M_c(n)] - \E[M_{c,S}(n)]$ evaluated numerically from Eq.\ (\ref{exact_E_3}) for Pareto and stable jump distributions, as a function of $n$ (in log-scale), corroborating the asymptotic behavior in Eq.~(\ref{general3}).}\label{fig_pareto}
\end{figure}

Figure\ \ref{fig_pareto} shows a numerical check of the generalization of our results beyond stable jump distributions. Here, it is a L\'evy flight with a Pareto jump distribution
\begin{eqnarray}\label{Pareto}
f(\eta) = 
\begin{cases}
&\dfrac{b^\mu}{2 \mu} |\eta|^{-1-\mu} \;, \; |\eta| \geq b \\
& 0 \;, |\eta| < b \;.
\end{cases}
\end{eqnarray}
All the parameters are the same as for the stable distribution considered in Fig. \ref{Fig_mu12}. Namely, $\mu=3/2$, $c=-1/2$, and $a=1$ [which corresponds to $b = (2 \pi)^{-1/3}$ in Eq.\ (\ref{Pareto})]. Fig.\ \ref{fig_pareto}a) is the counterpart of Fig.\ \ref{Fig_mu12}b). The comparison of these two figures clearly suggest the same leading behavior, $|c| C/(2-\mu)\,n^{2-\mu}$, and logarithmic correction, $|c|C_2\ln n$, but the next constant term is different. Indeed, the squares do have the same tendency as for the stable jump distribution to align parallel to the blue dotted line, but not at the same location [above the line in Fig.\ \ref{fig_pareto}a), below it in Fig.\ \ref{Fig_mu12}b)]. This corresponds to a different value of the constant term $\kappa_c$, in full agreement with our result in Sec.\ \ref{sec5.2.2} that the asymptotic expansions must be the same up to the constant term which is different. In Fig.\ \ref{fig_pareto}b), we show a plot of the difference $\E[M_c(n)] - \E[M_{c,S}(n)]$, evaluated numerically from the exact Eq.\ (\ref{exact_E_3}) for Pareto and stable jump distributions, as a function of $n$. Again, the results indicates that $\E[M_c(n)] - \E[M_{c,S}(n)]$ goes to a constant when $n$ is large, in agreement with Eq.\ (\ref{general3}). Note that the numerical evaluation of the integral in Eq.\ (\ref{general2}) to be used on the right-hand side of Eq.\ (\ref{general3}) requires a careful numerical analysis.

%
%
\section{Conclusion}\label{sec7}
In this paper, we have studied the asymptotic large $n$ behavior of the expected maximum $\E[M_c(n)]$ of an $n$-step random walk/L\'evy flight in the presence of a non-zero drift $c$. First, we have generalized the Hopf-Ivanov formula to the case of discrete time RWs with a drift [Eq.\ (\ref{eq2.1.5})], from which we have derived the Pollaczek-Spitzer formula for the PDF of the maximum in a new, convenient, factorized form [see Eq.\ (\ref{eq3.5})]. Performing then a careful analysis of the generating function of the expected maximum deduced from this version of the Pollaczek-Spitzer formula [Eq.\ (\ref{eq3.6b})], we have obtained a variety of interesting results that we now summarize.

In all cases, we have been able to determine the full asymptotic expansion of $\E[M_c(n)]$ comprising all the terms surviving the large $n$ limit, explicitly. For a nonzero drift $c\ne 0$ and $\nu\ge 2$ in Eq.\ (\ref{eq1.3}), we have found that this asymptotic expansion does {\it not} depend on the jump distribution (in the considered class, with same parameters $\mu$ and $a$ in Eq.\ (\ref{eq1.3}), and same drift $c$) except the smallest, constant, term which does depend on the jump PDF. The same result holds in the driftless case $c=0$, but for a smaller class of jump PDF defined by $\nu>\mu +1$ in Eq.\ (\ref{eq1.3}). For a negative drift $c<0$ and L\'evy index $\mu =2$, the expected maximum approaches a constant as $n$ gets large, while for $1<\mu < 2$, it grows as a power law $\sim n^{2-\mu}$. It is interesting to mention that, in both cases, it can be shown from the Pollaczek-Spitzer formula\ (\ref{eq3.5}) that the distribution of the maximum has a well defined limiting distribution $p_{-|c|}(\m)=\lim_{n\to +\infty}p_{-|c|}(\m,n)$ with $p_{-|c|}(\m)\sim \m^{-\mu}$ ($\m\to +\infty$) for $1<\mu <2$ (not discussed in this paper). We have also found that the asymptotic expansion of $\E[M_c(n)]$ for a positive drift $c>0$ is related to the one with a negative drift by the general identity $\E[M_c(n)]=cn+\E[M_{-c}(n)]$, which we have explained very simply by symmetry arguments (see Fig. \ref{fig_identity}). Another interesting finding is that the $c \to 0$ and $n \to \infty$ limits of $\E[M_c(n)]$ do not commute. This led us to derive a scaling form interpolating smoothly between the leading behaviors of $\E[M_c(n)]$ in the two limiting cases `$c \to 0$ then $n \to \infty$' and `$n \to \infty$ then $c \to 0$' [Eqs.\ (\ref{cross_3}) and\ (\ref{cross_4})].

This paper was devoted to developing the formalism and analytical tools to extract the asymptotic behaviors of the expected maximum. Besides the generalized Hopf-Ivanov formula which provides a useful tool to study other observables associated to extremes, order, and record statistics of biased RWs, we expect that our results will have a number of interesting applications in other problems. These include, for instance, the mean perimeter of
the convex hull of a $2D$-random walk process \cite{GLM17} and the expected flux in the Schmoluchovski capture problem \cite{Smoluchowsky,MCZ06}. These two problems have been investigated in the absence of a drift and we expect that our results will be useful to generalize these studies in the presence of a nonzero drift. 
%
%
\acknowledgements 
We thank Alain Comtet for useful discussions.

\appendix
\section{Derivation of the expression of $\bm{\phi^{(c)}(\lambda ,s)}$}\label{app1}
From
\begin{equation}\label{eqA1.1a}
p_{>0}^{(c)}(x_2,n\vert x_1,0)=\int_0^{+\infty}f(x_2-y-c)\, p_{>0}^{(c)}(y,n-1\vert x_1,0)\, dy,
\end{equation}
and
\begin{equation}\label{eqA1.1b}
p_{>0}^{(c)}(x_2,n\vert x_1,0)=\int_0^{+\infty}p_{>0}^{(c)}(x_2,n-1\vert y,0)\, f(y-x_1-c)\, dy,
\end{equation}
one respectively gets
\begin{equation}\label{eqA1.2a}
G_{>0}^{(c)}(x_2,x_1,s)-s\int_0^{+\infty}f(x_2-y-c)\, G_{>0}^{(c)}(y,x_1,s)\, dy =\delta(x_2-x_1),
\end{equation}
and
\begin{equation}\label{eqA1.2b}
G_{>0}^{(c)}(x_2,x_1,s)-s\int_0^{+\infty}G_{>0}^{(c)}(x_2,y,s)\, f(y-x_1-c)\, dy =\delta(x_2-x_1),
\end{equation}
where we have used $p_{>0}^{(c)}(x_2,0\vert x_1,0)=\delta(x_2-x_1)$. Write $u^{(c)}(x,s)=G_{>0}^{(c)}(x,0,s)$. For $x_1=0$ and $x_2=x$ the equation\ (\ref{eqA1.2a}) reads
\begin{equation}\label{eqA1.3}
u^{(c)}(x,s)-s\int_0^{+\infty}f(x-y-c)u^{(c)}(y,s)\, dy =\delta(x).
\end{equation}
Here and in the following, $\delta(x)$ is to be understood as $\delta(x-0^+)$. Deriving\ (\ref{eqA1.3}) with respect to $s$ and using\ (\ref{eqA1.3}) in the result, one has
\begin{equation}\label{eqA1.4}
\frac{\partial u^{(c)}(x,s)}{\partial s}
-s\int_0^{+\infty}f(x-y-c)\frac{\partial u(y,s)}{\partial s}\, dy =
\frac{u(x,s)-\delta(x)}{s}.
\end{equation}
From\ (\ref{eqA1.4}) and\ (\ref{eqA1.2b}) it follows
\begin{equation}\label{eqA1.5}
s\frac{\partial u^{(c)}(x,s)}{\partial s}=\int_0^{+\infty}G_{>0}^{(c)}(x,y,s)
\lbrack u^{(c)}(y,s)-\delta(y)\rbrack\, dy,
\end{equation}
and, by Laplace transforming\ (\ref{eqA1.5}) with respect to $x$,
\begin{eqnarray}\label{eqA1.6}
s\frac{\partial \phi^{(c)}(\lambda ,s)}{\partial s}&=&\int_0^{+\infty}\tilde{G}_{>0}^{(c)}(\lambda ,y,s)
\lbrack u^{(c)}(y,s)-\delta(y)\rbrack\, dy \nonumber \\
&=&\int_0^{+\infty}\tilde{G}_{>0}^{(c)}(\lambda ,y,s)u^{(c)}(y,s)\, dy -\tilde{G}_{>0}^{(c)}(\lambda ,0,s).
\end{eqnarray}
Here, $\lambda$ is the Laplace variable and
\begin{eqnarray}\label{eqA1.7}
\tilde{G}_{>0}^{(c)}(\lambda ,y,s)&=&\int_0^{+\infty}G_{>0}^{(c)}(x,y,s){\rm e}^{-\lambda x}dx \nonumber \\
&=&\int_0^{+\infty}\int_0^{\min(x,y)}
u^{(c)}(x-z,s)u^{(-c)}(y-z,s){\rm e}^{-\lambda x}\, dz\, dx \nonumber \\
&=&\int_0^{y}u^{(-c)}(y-z,s)\int_z^{+\infty}u^{(c)}(x-z,s){\rm e}^{-\lambda x}\, dx\, dz \nonumber \\
&=&\phi^{(c)}(\lambda ,s)\int_0^{y}u^{(-c)}(y-z,s){\rm e}^{-\lambda z}dz,
\end{eqnarray}
where we have used\ (\ref{eq2.1.4}). Note that, by the singular behavior of $u^{(c)}(x,s)$ near $x=0$ [see\ (\ref{eqA1.3})], Equation\ (\ref{eqA1.7}) for $y=0$ reduces to $\tilde{G}_{>0}^{(c)}(\lambda ,0,s)=\phi^{(c)}(\lambda ,s)$, as it should be. Now, substituting for $\tilde{G}_{>0}^{(c)}(\lambda ,y,s)$ on the right-hand side of\ (\ref{eqA1.6}) its expression from\ (\ref{eqA1.7}), one finds
\begin{eqnarray}\label{eqA1.8}
s\frac{\partial \phi^{(c)}(\lambda ,s)}{\partial s}&=&\phi^{(c)}(\lambda ,s)\left\lbrack
\int_0^{+\infty}\int_0^y u^{(-c)}(y-z,s)u^{(c)}(y,s){\rm e}^{-\lambda z}\, dz\, dy -1\right\rbrack  \\
&=&\phi^{(c)}(\lambda ,s)\left\lbrack\int_0^{+\infty}\left(\int_z^{+\infty}
u^{(-c)}(y-z,s)u^{(c)}(y,s)\, dy\right) {\rm e}^{-\lambda z}dz -1\right\rbrack \nonumber \\
&=&\phi^{(c)}(\lambda ,s)\left\lbrack\int_0^{+\infty}\left(\int_0^{+\infty}u^{(-c)}(t,s)u^{(c)}(t+z,s)\, dt\right)
{\rm e}^{-\lambda z}dz -1\right\rbrack \nonumber \\
&=&\phi^{(c)}(\lambda ,s)\int_0^{+\infty}\left(\int_0^{+\infty}u^{(-c)}(t,s)u^{(c)}(t+z,s)\, dt -\delta(z)\right)
{\rm e}^{-\lambda z}dz, \nonumber
\end{eqnarray}
where $t=y-z$. To proceed we take $x_1=X$ and $x_2=X+x$, $x\ge 0$, in Eq.\ (\ref{eq2.1.4}). Then, we make the change of variable $y=X-t$, and let $X\rightarrow +\infty$. Using $\lim_{X\rightarrow +\infty}G_{>0}^{(c)}(X+x,X,s)=G_{>-\infty}^{(c)}(x,0,s)$, one finds
\begin{equation}\label{eqA1.9}
G_{>-\infty}^{(c)}(x,0,s)=\int_0^{+\infty}
u^{(-c)}(t,s)u^{(c)}(t+x,s)\, dt,
\end{equation}
and Eq.\ (\ref{eqA1.8}) reduces to
\begin{equation}\label{eqA1.10}
s\frac{\partial\ln\phi^{(c)}(\lambda ,s)}{\partial s}=\tilde{\Xi}^{(c)}(\lambda ,s),
\end{equation}
where $\tilde{\Xi}^{(c)}(\lambda ,s)$ is the Laplace transform of $\Xi^{(c)}(x,s)=G_{>-\infty}^{(c)}(x,0,s)-\delta(x)$ with respect to $x$. The equation for $G_{>-\infty}^{(c)}(x,0,s)$ is similar to\ (\ref{eqA1.3}) with a full-space integration, which is readily solved by Fourier transform. One gets
\begin{equation}\label{eqA1.11}
G_{>-\infty}^{(c)}(x,0,s)=\frac{1}{2\pi}\int_{-\infty}^{+\infty}
\frac{\exp(-ikx)}{1-s\hat{f}(k){\rm e}^{ikc}}\, dk,
\end{equation}
hence
\begin{equation}\label{eqA1.12}
\Xi^{(c)}(x,s)=\frac{1}{2\pi}\int_{-\infty}^{+\infty}\frac{s\hat{f}(k){\rm e}^{ikc}}{1-s\hat{f}(k){\rm e}^{ikc}}
{\rm e}^{-ikx}\, dk,
\end{equation}
and, by Fubini's theorem,
\begin{eqnarray}\label{eqA1.13}
\tilde{\Xi}^{(c)}(\lambda ,s)&=&\frac{1}{2\pi}\int_0^{+\infty}{\rm e}^{-\lambda x}\left(\int_{-\infty}^{+\infty}
\frac{s\hat{f}(k){\rm e}^{ikc}}{1-s\hat{f}(k){\rm e}^{ikc}}{\rm e}^{-ikx}\, dk\right) dx \nonumber \\
&=&\frac{1}{2\pi}\int_{-\infty}^{+\infty}\frac{s\hat{f}(k){\rm e}^{ikc}}{1-s\hat{f}(k){\rm e}^{ikc}}
\left(\int_0^{+\infty}{\rm e}^{-(\lambda+ik)x}\, dx\right) dk \nonumber \\
&=&\frac{1}{2\pi}\int_{-\infty}^{+\infty}
\frac{1}{\lambda +ik}\left\lbrack\frac{s\hat{f}(k){\rm e}^{ikc}}{1-s\hat{f}(k){\rm e}^{ikc}}\right\rbrack dk.
\end{eqnarray}
Substituting\ (\ref{eqA1.13}) for $\tilde{\Xi}^{(c)}(\lambda ,s)$ on the right-hand side of\ (\ref{eqA1.10}) yields
\begin{equation}\label{eqA1.14}
\frac{\partial\ln\phi^{(c)}(\lambda ,s)}{\partial s}=\frac{1}{2\pi}\int_{-\infty}^{+\infty}
\frac{1}{\lambda +ik}\left\lbrack\frac{\hat{f}(k){\rm e}^{ikc}}{1-s\hat{f}(k){\rm e}^{ikc}}\right\rbrack dk.
\end{equation}
Finally, integrating\ (\ref{eqA1.14}) with $\phi^{(c)}(\lambda ,0)=1$ [which follows from $u^{(c)}(x,0)=\delta(x)$], one obtains the equation\ (\ref{eq2.1.6}):
\begin{equation*}
\phi^{(c)}(\lambda ,s)=\exp\left(-\frac{1}{2\pi}\int_{-\infty}^{+\infty}
\frac{\ln\lbrack 1-s\hat{f}(k){\rm e}^{ikc}\rbrack}{\lambda +ik}\, dk\right).
\end{equation*}
%
%
\section{Derivation of the Sparre Andersen formula for a random walk with a drift}\label{App_SA}
In this Appendix, using the generalized Hopf-Ivanov formula (\ref{eq2.1.5}) and the representation of $\phi^{(c)}(\lambda ,s)$ in Eqs. (\ref{eq2.2.2b}) and (\ref{eq2.2.3b}), we derive the generalized Sparre Andersen theorem for the survival probability of the walk in the presence of a drift. We start by recalling the Sparre Andersen theorem. Let $Q^{(c)}(n)$ denote the persistence, i.e., the probability that the random walk in Eq. (\ref{eq1.1}), starting at $x_0=0$ stays above its initial value $x_0=0$ up to step $n$. A beautiful combinatorial identity, first derived by Sparre Andersen in \cite{SA1954}, reads
\begin{equation}\label{SA0}
\sum_{n=0}^{+\infty}Q^{(c)}(n)s^n=\exp\left\lbrack\sum_{n=1}^{+\infty} \frac{s^n}{n} {\rm Prob}(x_n \geq 0) \right\rbrack
\end{equation}
where $x_n$ evolves by Eq. (\ref{eq1.1}) with a non-zero $c$. This identity is nontrivial as it relates a history dependent quantity $Q^{(c)}(n)$ on the l.h.s. with a temporally local quantity ${\rm Prob}(x_n \geq 0)$. Writing $x_n = y_n + c\,n$, we see that the sequence $y_n$ corresponds to a random walk without a drift, evolving as
\bea\label{evol_yn}
y_n = y_{n-1} + \eta_n \;,
\eea
and hence
\bea\label{yn_1}
{\rm Prob}(x_n \geq 0) = {\rm Prob}(y_n \geq -c\,n) = \int_{-c\,n}^{+\infty} P_n(y) \,dy
\eea
where $P_n(y)$ is the PDF of the driftless $y$-process. This can be easily computed and is given by
\bea\label{Pny_app}
P_n(y) = \frac{1}{2\pi} \int_{-\infty}^{+\infty} \left[\hat f(k)\right]^n e^{-ik\,y} dk \;.
\eea
Therefore, the Sparre Andersen theorem can be re-expressed as
\begin{equation}\label{SA1}
\sum_{n=0}^{+\infty}Q^{(c)}(n)s^n=\exp\left\lbrack
\sum_{n=1}^{+\infty}\frac{s^n}{n}\int_{-cn}^{+\infty}P_n(y)\, dy\right\rbrack ,
\end{equation}
where $P_n(y)$ is given in (\ref{Pny_app}). Below we prove this identity (\ref{SA1}), starting from the generalized Hopf-Ivanov formula in (\ref{eq2.1.5}) and using the alternative expression of $\phi^{(c)}(\lambda,s)$ derived in Eqs.\ (\ref{eq2.2.2b}) and\ (\ref{eq2.2.3b}). 

We start with the generalized Hopf-Ivanov formula in (\ref{eq2.1.5})  where we set $\lambda_1\rightarrow +\infty$ and $\lambda_2=0$. One finds
\begin{eqnarray}\label{SA2}
\sum_{n\ge 0} s^n \, \int_0^{+\infty} p^{(c)}_{>0}(x,n\vert 0,0)\, dx =\phi^{(c)}(0,s) \;.
\end{eqnarray}
Note however that the integral $\int_0^{+\infty} p^{(c)}_{>0}(x,n\vert 0,0)\, dx$ is just the probability that the process stays positive up to step $n$, and hence is identically equal to $Q^{(c)}(n)$. Thus we get
\begin{eqnarray}\label{SA2_bis}
\sum_{n\ge 0} s^n \, Q^{(c)}(n) = \phi^{(c)}(0,s) \;.
\end{eqnarray}
Now we evaluate $ \phi^{(c)}(0,s)$ using Eqs.\ (\ref{eq2.2.2b}) and\ (\ref{eq2.2.3b}). This gives
\begin{eqnarray}\label{SA3}
\sum_{n\ge 0} s^n \, Q^{(c)}(n) = \phi^{(c)}(0,s) = \exp\left\lbrack
\sum_{n=1}^{+\infty}\frac{s^n}{n}\int_0^{+\infty}
{\cal F}_{\mu ,n}\left(x-n^{1-1/\mu}c\right)\, dx\right\rbrack .
\end{eqnarray}
Comparing Eq.\ (\ref{eq2.2.3a}) and Eq.~(\ref{Pny_app}) we see that
\bea\label{SA3_bis}
{\cal F}_{\mu ,n}(x)=n^{1/\mu}P_n(n^{1/\mu}x) \;.
\eea
Substituting (\ref{SA3_bis}) on the right-hand side of (\ref{SA3}), and performing a change of variable $x=n^{-1/\mu}y$, we get 
\begin{eqnarray*}
\sum_{n=0}^{+\infty}Q^{(c)}(n)s^n&=&\exp\left\lbrack
\sum_{n=1}^{+\infty}\frac{s^n}{n}\int_{0}^{+\infty}P_n(y-cn)\, dy\right\rbrack \nonumber \\
&=&\exp\left\lbrack
\sum_{n=1}^{+\infty}\frac{s^n}{n}\int_{-cn}^{+\infty}P_n(y)\, dy\right\rbrack ,
\end{eqnarray*}
which thus proves the Sparre Andersen identity in (\ref{SA1}).
%
%
\section{Connection with Spitzer's formula (Theorem 3 in \cite{Spi57})}\label{App_Spitzer}
In this Appendix, we show the connection between our formula (\ref{eq3.5}) and the formula obtained by Spitzer, see Theorem 3 given in Eq. (3.1) of Ref. \cite{Spi57}, which reads, in our notation,
\begin{eqnarray}\label{Spit1}
&&\sum_{n\ge 0}\E\left({\rm e}^{-\lambda M_c(n)}\right)s^n = \nonumber \\
&&\frac{1}{1-s} \exp{\left(\frac{1}{2\pi} \int_0^s d\tau \int_{-\infty}^{+\infty} \frac{\lambda}{k(k-i\lambda)}  \frac{\hat f(k) \e^{i k c} - 1}{(1-\tau)(1 - \tau \hat f(k)\,\e^{i k c})} dk\right)} \;.
\end{eqnarray}
In Eq. (\ref{Spit1}), the integral over $k$ must be understood as
\begin{eqnarray}\label{regularization}
&&\int_0^s d\tau \int_{-\infty}^{+\infty} \frac{\lambda}{k(k-i\lambda)}  \frac{\hat f(k) \e^{i k c} - 1}{(1-\tau)(1 - \tau \hat f(k)\,\e^{i k c})} dk \\
&=& \lim_{\epsilon \to 0^+} \int_0^s d\tau \int_{-\infty}^{+\infty} \frac{\lambda-\epsilon}{(k- i \epsilon)(k-i\lambda)}  \frac{\hat f(k) \e^{i k c} - 1}{(1-\tau)(1 - \tau \hat f(k)\,\e^{i k c})} dk \;,
\end{eqnarray}
(see Eqs. (3.4) and below of Ref. \cite{Spi57}). Performing the integral over $\tau$ in Eq. (\ref{regularization}) one gets
\begin{eqnarray}\label{Spit2}
&&\int_0^s d\tau \int_{-\infty}^{+\infty} \frac{\lambda}{k(k-i\lambda)}  \frac{\hat f(k) \e^{i k c} - 1}{(1-\tau)(1 - \tau \hat f(k)\,\e^{i k c})} dk \nonumber \\
&& =  \lim_{\epsilon \to 0^+}\int_{-\infty}^{+\infty} \frac{\lambda-\epsilon}{(k- i \epsilon)(k-i\lambda)} \left[\ln(1-s) -\ln(1 - s \hat f(k)\e^{i k c})\right] dk \nonumber \\
&& = - \lim_{\epsilon \to 0^+}\int_{-\infty}^{+\infty} \frac{\lambda-\epsilon}{(k- i \epsilon)(k-i\lambda)} \ln(1 - s \hat f(k)\e^{i k c})\, dk \;,
\end{eqnarray}
where the last line follows from
\begin{eqnarray}
\int_{-\infty}^{+\infty} \frac{dk}{(k-i \epsilon)(k - i \lambda)} = 0
\end{eqnarray}
for $\epsilon, \lambda > 0$. Using then
\begin{eqnarray}
\frac{\lambda-\epsilon}{(k-i\epsilon)(k-i \lambda)} = \frac{1}{\lambda + ik} - \frac{1}{\epsilon + i k}
\end{eqnarray}
in Eq. (\ref{Spit2}), one finds that the formula obtained by Spitzer in (\ref{Spit1}) can be rewritten as
\begin{eqnarray}\label{Spit3_0}
\sum_{n\ge 0}\E\left({\rm e}^{-\lambda M_c(n)}\right)s^n
&=& \frac{1}{1-s} \exp{\left(-\int_{-\infty}^{+\infty} \frac{dk}{\lambda + ik}  \ln(1 - s \hat f(k)\e^{i k c}) \right)} \nonumber  \\
&\times&\lim_{\epsilon \to 0^+}
\exp{\left(\int_{-\infty}^{+\infty} \frac{dk}{\epsilon + ik}  \ln(1 - s \hat f(k)\e^{i k c}) \right)},
\end{eqnarray}
or, in a more compact form, using in\ (\ref{Spit3_0}) the expression of $\phi^{(c)}(\lambda,s)$ given in Eq.~(\ref{eq2.1.6}),
\begin{equation}\label{Spit3}
\sum_{n\ge 0}\E\left({\rm e}^{-\lambda M_c(n)}\right)s^n =
\frac{1}{1-s} \frac{\phi^{(c)}(\lambda,s)}{\phi^{(c)}(0,s)}.
\end{equation}
To show that this formula (\ref{Spit3}), derived from the Spitzer's result in (\ref{Spit1}), coincides with our result given in Eq. (\ref{eq3.5}), we need to establish the following identity
\begin{eqnarray}\label{identity}
\phi^{(c)}(0,s) \phi^{(-c)}(0,s) = \lim_{\epsilon \to 0} \phi^{(c)}(\epsilon,s) \phi^{(-c)}(\epsilon,s) = \frac{1}{1-s} \;.
\end{eqnarray}
To this end, we write the product $\phi^{(c)}(\epsilon,s) \phi^{(-c)}(\epsilon,s)$ explicitly by using the expression given in (\ref{eq2.1.6}), i.e.,
\begin{eqnarray}\label{identity2}
&&\phi^{(c)}(\epsilon,s) \phi^{(-c)}(\epsilon,s) = \nonumber \\
&&\exp\left(-\frac{1}{2\pi} \int_{-\infty}^{+\infty} \frac{\ln(1 - s\hat f(k) e^{ikc})}{\epsilon + i k}\, dk -\frac{1}{2\pi} \int_{-\infty}^{+\infty} \frac{\ln(1 - s\hat f(k) e^{-ikc})}{\epsilon + i k}\, dk \right) \;.
\end{eqnarray}
In the second integral we perform the change of variable $k \to -k$ and use that $\hat f(k) = \hat f(-k)$ to obtain
\begin{eqnarray}\label{identity3}
\phi^{(c)}(\epsilon,s) \phi^{(-c)}(\epsilon,s) = \exp{\left(-\frac{1}{\pi} \int_{-\infty}^{+\infty} dk \frac{\epsilon}{\epsilon^2 + k^2} \ln(1 - s \hat f(k)e^{i k c})\right)}
\end{eqnarray}
It remains to make the change of variable $k = \epsilon q$ on the right-hand side of\ (\ref{identity3}) and let $\epsilon \to 0$ to get (using $\hat f(0) = 1$)
\begin{eqnarray}
\lim_{\epsilon \to 0^+} \phi^{(c)}(\epsilon,s) \phi^{(-c)}(\epsilon,s) = \exp{\left(-\frac{1}{\pi} \int_{-\infty}^{+\infty} \frac{dq}{q^2+1} \ln (1-s) \right)} = \frac{1}{1-s} \;,
\end{eqnarray}
which demonstrates the identity (\ref{identity}). Finally, using this identity\ (\ref{identity}) in Eq.\ (\ref{Spit3}) establishes that the Spitzer's formula (\ref{Spit1}) coincides with our result in Eq. (\ref{eq3.5}). 

\begin{thebibliography}{1}
%

\bibitem{bouchaud_biroli}
Biroli, G., Bouchaud, J.-Ph., Potters, M.: Extreme value problems in Random Matrix Theory and other disordered systems, J. Stat. Mech., P07019 (2007).

\bibitem{satya_review}
Majumdar, S. N.: Universal first-passage properties of discrete-time random walks and L\'evy flights on a line: Statistics of the global maximum and records, Physica A, {\bf 389}, 4299 (2010).

\bibitem{third}
Majumdar, S. N., Schehr, G.: Top eigenvalue of a random matrix: large deviations and third order phase transition, J. Stat. Mech. P01012 (2014).

\bibitem{review_record}
Godr\`eche, C., Majumdar, S. N., Schehr, G.: Record statistics of a strongly correlated time series: random walks and L\'evy flights, J. Phys. A: Math. Theor. {\bf 50}, 333001 (2017)

\bibitem{gumbel}
Gumbel, E. J., {\it Statistics of Extremes}, Columbia University Press, (1958).

\bibitem{Pal_Satya}
Majumdar, S. N., Pal, A.: Extreme value statistics of correlated random variables, arXiv preprint arXiv:1406.6768 (2014).

\bibitem{Spi57}
Spitzer, F.: The Wiener-Hopf equation whose kernel is a probability density, Duke Math. J. {\bf 24}, 327 (1957).

\bibitem{AS2005} Comtet, A., Majumdar, S. N.: Precise asymptotics for a random walker's maximum, J. Stat. Mech. P06013 (2005).

\bibitem{GLM17} Grebenkov, D. S., Lanoisel\'ee, Y., Majumdar, S. N.: Mean perimeter and mean area of the convex hull over planar random walks, J. Stat. Mech. P103203 (2017). 

\bibitem{flajolet}
Coffman, E. G., Flajolet, Ph., Flato, L., Hofri, M.: The maximum of random walk and its application to rectangle packing, Probab. Eng. Inform. Sc. {\bf 12}, 373 (1998).

\bibitem{Hopf}
Hopf, E., {\it Mathematical problems of radiative equilibrium}, Cambridge University Press, Cambridge (1934).

\bibitem{Ivanov} Ivanov, V. V.: Resolvent method: exact solutions of half-space transport problems by elementary means, Astron. Astrophys. {\bf 286}, 328 (1994).
 
\bibitem{Yor_book}
Revuz, D., Yor, M. {\it Continuous martingales and Brownian motion}, (Vol. 293), Springer Science \& Business Media (2013).

\bibitem{pers_review} Bray, A. J., Majumdar, S. N., Schehr, G.: Persistence and first-passage properties in non-equilibrium systems, Adv. Phys. {\bf 62}, 225 (2013).

\bibitem{MC02} Majumdar, S. N., Comtet, A.: Exact asymptotic results for persistence in the Sinai model with arbitrary drift, Phys. Rev. E {\bf 66}, 061105 (2002).

\bibitem{MMS2014} Majumdar, S. N., Mounaix, Ph., Schehr, G.: On the gap and time interval between the first two maxima of long random walks, J. Stat. Mech. P09013 (2014).

\bibitem{SA1954} Sparre Andersen, E.: On the fluctuations of sums of random variables II, Math. Scand. {\bf 2}, 195 (1954).

\bibitem{Hen} Henrici, P.: Applied and Computational Complex Analysis. Vol. 2, Wiley Classics Library, John Wiley, New York, 1991. Theorem 11.10b: ``Theorem of Darboux".

\bibitem{dembo} Dembo, A.: private communication.

\bibitem{Ell} Ellis, R. S.: The theory of large deviations: from Boltzmann's 1877 calculation to equilibrium macrostates in 2D turbulence, Physica D {\bf 133}, 106 (1999).
%
\bibitem{Nag} Nagaev, A. V.: Integral limit theorems for large deviations when Cram\'er's condition is not fulfilled I \& II, Theory Probab. Appl. {\bf 14}, No 1, 51 \& No 2, 193 (1969).

\bibitem{DDS2008} Denisov, D., Dieker, A. B., Shneer, V.: Large deviation for random walks under subexponentiality: the big-jump domain, Ann. Probab. {\bf 36}, 1946 (2008).

\bibitem{FS} Flajolet, Ph., Sedgewick, R.: Analyic Combinatorics, Cambridge University Press, Cambridge, 2009.
%
\bibitem{Cram} Cram\'er, H.: Sur un nouveau th\'eor\`eme limite dans la th\'eorie des probabilit\'es. In {\it Colloque consacr\'e \`a la th\'eorie des probabilit\'es}, volume 3, pp. 2-23, Hermann, Paris, 1938.

\bibitem{Smoluchowsky} M. von Smoluchowski, Phys. Z.
{\bf 17}, 557-571, 585-599 (1916). 

\bibitem{MCZ06} S. N. Majumdar, A. Comtet, R. M. Ziff, J. Stat. Phys.
{\bf 122}, 833 (2006).

%
%
%
%
%

%

%
%



















\end{thebibliography}
\end{document}